\documentclass[acmlarge, nonacm]{acmart}

\AtBeginDocument{%
  \providecommand\BibTeX{{%
    \normalfont B\kern-0.5em{\scshape i\kern-0.25em b}\kern-0.8em\TeX}}}

\usepackage{xspace}
\newcommand{\nickName}{\textsc{Company X}\xspace}
\usepackage{subfigure, caption, graphicx}
\usepackage{multirow}
\begin{document}

\title{Understanding Diffusion of Recurrent Innovations}

\author{Fuqi Lin}
\email{linfuqi@pku.edu.cn}
\affiliation{%
  \institution{Peking University}
  \streetaddress{No.5 Yiheyuan Road}
  \city{Beijing}
  \country{China}
  \postcode{100871}
}

\makeatletter
\let\@authorsaddresses\@empty
\makeatother

\begin{abstract}
  The diffusion of innovations theory has been studied for years. Previous research efforts mainly focus on key elements, adopter categories, and the process of innovation diffusion. However, most of them only consider single innovations. With the development of modern technology, recurrent innovations gradually come into vogue. In order to reveal the characteristics of recurrent innovations, we present the first large-scale analysis of the adoption of recurrent innovations in the context of mobile app updates. Our analysis reveals the adoption behavior and new adopter categories of recurrent innovations as well as the features that have impact on the process of adoption.

\end{abstract}

\maketitle
\section{Introduction}\label{sec:introduction}
Modern technology innovation functions in a recurrent nature. Smartphones did not take over in one shot: many generations of mobile phones rose and replaced each other in the past decades, and many have vanished from memory, leaving version numbers in the ash. Car models update every couple of years, delivering new versions of autonomous driving. New neural networks improve upon older ones faster than new updates of the TikTok app. The decision of customers have switched from \textit{should I get a typewriter} to \textit{should I get the newest Macbook pro}. Understanding how users adopt recurrent, evolving innovations is crucial for innovators, stakeholders, disseminators, and retailers. 

Adoption of innovations has been studied for decades since Everett Rogers' classic book \textit{Diffusion of Innovations}~\cite{rogers2010diffusion}, first published in 1962. As defined in Rogers' theory, an innovation is \textit{an idea, practice, or object perceived as new by an individual or other unit of adoption}, and adopters of an innovation are categorized into \textit{innovators}, \textit{early adopters}, \textit{early majority}, \textit{late majority}, and \textit{laggards} \cite{rogers2010diffusion}. The theory of the diffusion of innovations and its variations have been widely applied to multiple disciplines. such as medical sociology~\cite{coleman1957the}, cultural anthropology~\cite{barnett1963innovation}, industrial economics~\cite{mansfield1985rapidly}, health care~\cite{dearing2018diffusion}, as well as mobile technologies~\cite{liang2007adoption, kauffman2005international} and apps~\cite{nickerson2014mobile, east2015mental, grinko2019adoption, yujuico2015considerations}. 

However, most theories and empirical analyses have been developed upon one-time adoption of single innovations, and little work has been done regarding the adoption of recurrent innovations.  This is possibly due to the nature of traditional innovations, but more likely because of the difficulty of collecting relevant data. Indeed, understanding recurrent innovations requires tracking the adopters of different versions of innovations over a long period of time, which has not been feasible in most domains. 

\begin{figure*}[htb]
    \centering
    \begin{center}
        \subfigure[iOS\label{fig:total-line-mom}]
        {\includegraphics[width=0.48\linewidth]{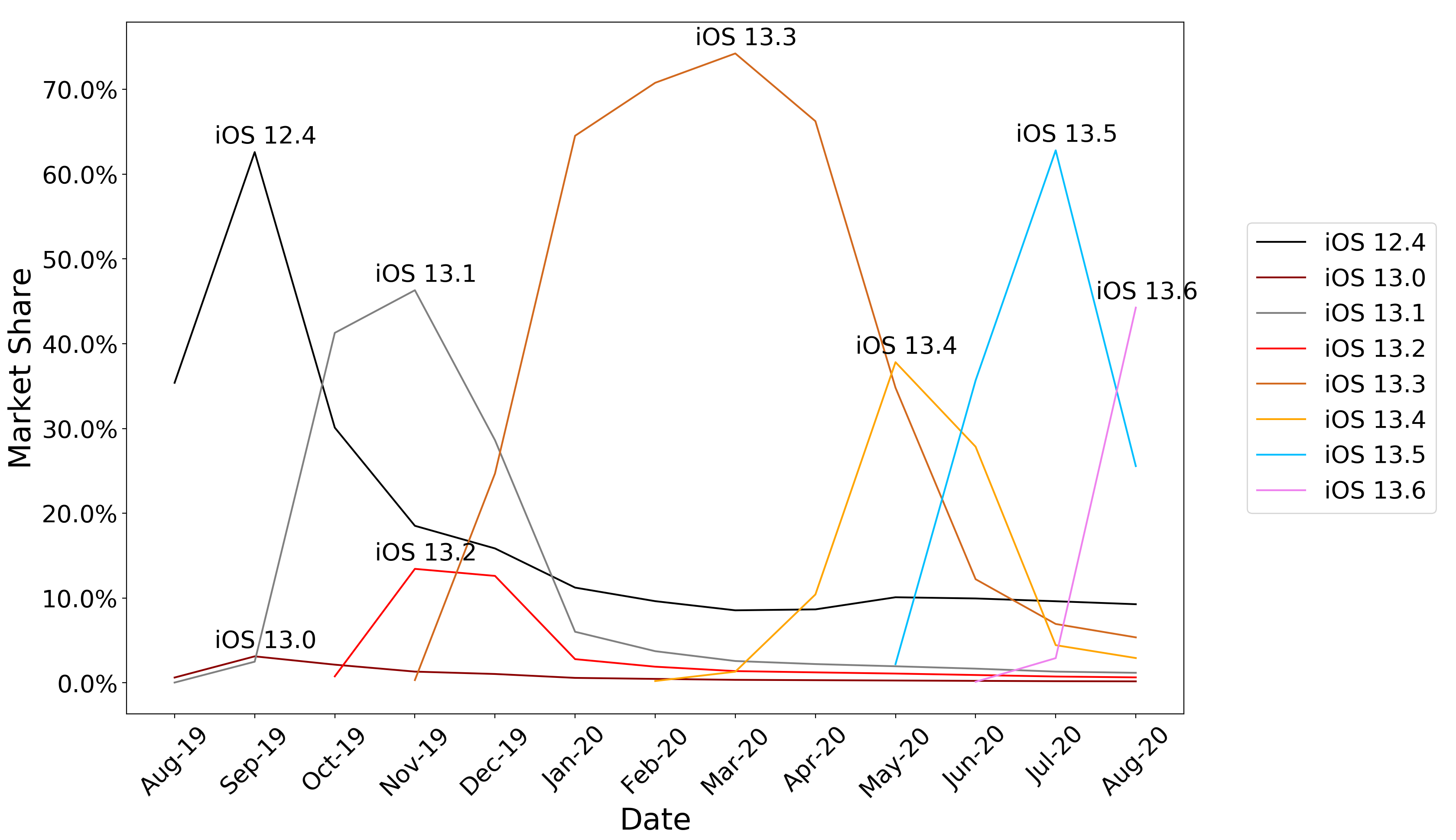}}
        \subfigure[Chrome\label{fig:total-stack-mom}]
        {\includegraphics[width=0.48\linewidth]{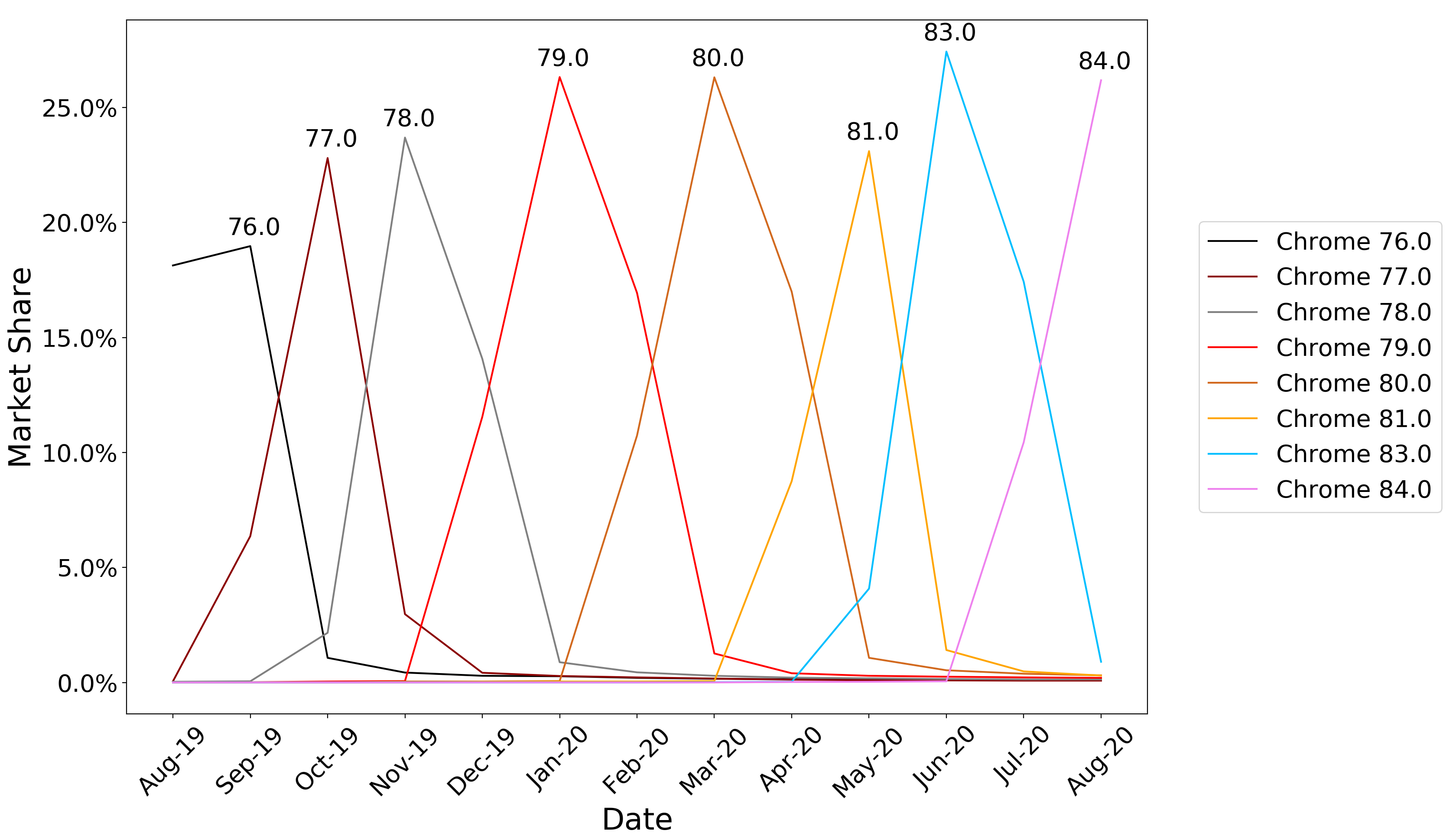}}
        \caption{Market share increases of old versions of iOS and Chrome slow down and reverse after new version releases.} \label{fig:market_share}
    \end{center}
\end{figure*}

It is therefore critical to verify whether these existing theories and findings still hold in the context of recurrent innovations. Specifically, the adoption behavior of a user can be largely influenced by whether or not they have adopted and liked earlier versions of the same technology; the diffusion process of an innovation can be affected by its subsequent updates.  As an example, a public dataset reveals the evolution of the market share of iOS,\footnote{https://gs.statcounter.com/os-version-market-share/ios/mobile-tablet/worldwide} the mobile operating system for Apple devices, and Chrome,\footnote{https://gs.statcounter.com/browser-version-market-share} a popular Web browser, both with multiple continuous versions. As is shown in Figure~\ref{fig:market_share}, the diffusion (or the increase of market share) of each version slows down or reverses after a newer version is released, demonstrating evidence of the possible interference between recurrent versions of innovations. Such a pattern already places a threat to existing conclusions about single innovations: according to the Rogers' theory, the market share of an innovation saturates but does not drop. What these data do not untangle are the adoption behaviors of individual users: whether there exist variations in the tendency of adoption, timeliness of adoption, and order of adoptions of new versions by different users, and if yes, whether these variations can be explained by the categorization of adopters in the classical diffusion of innovation theory. 

In this paper, we conduct the first systematic analysis of the diffusion of recurrent innovations, using a novel dataset that tracks how 17,124,831 Android users adopt 217,285 of Android applications collected by \nickName\footnote{Name of company masked for blind review}, a leading third-party data intelligence service provider.  This dataset provides a unique case scenario of recurrent innovations in the context of mobile apps. We find that the release of a new version of the same app clearly hinders the diffusion of its precedent versions. 
When adopting a new version of an app, there exists a significant difference between users who have not adopted a previous version of the app and those who have.  While the adoption curve of the former group (new adopters) comply with Rogers' theory, the latter group (recurrent adopters) presents considerably different patterns.  
In particular, we identify three novel categories of adopters of recurrent innovations, which are not covered in existing theories, namely the \textit{subscribers}, the \textit{preservers}, and the \textit{retro-adopters}.  These new categories account for a considerable proportion of adopters of an innovation, which has not been explored by previous efforts. Their interactions with precedent innovations present a clear difference from the rest of the users, clear enough for a machine learning model to distinguish them with decent accuracy.   

We further investigate whether a user's behavior of adopting a recurrent innovation is predictable. We find that using a group of features to represent the properties of the target app (the technology), the characteristics of the user (the adopter), and how the user interacts with previous versions of the app (recurrent innovations of the technology), off-the-shelf machine learning models are able to predict whether the user will adopt a new version of the app, and if yes, how soon the adoption will happen after the new version is released. The results reconfirm part of the existing theories about single innovations while once again reveal critical new insights about the factors that affect the adoption of a recurrent innovation. 

The contribution of this work can be summarized as follows:

\begin{itemize}
  \item To the best of our knowledge, this is the first large-scale analysis of the diffusion of recurrent innovations. Our work provides a novel perspective that augments the existing theories and models of the diffusion of single innovations.
  \item We base our analysis on a new longitudinal dataset tracking how tens of millions of users adopt 713,935 versions of mobile apps. This provides a unique application scenario and testbed to understand the diffusion of recurrent innovations. 
  \item We find significantly different patterns in the adoption curves of recurrent innovations that cannot be explained by existing theories, through which we identify new categories of adopters in the diffusion process. 
  \item We conduct a systematic predictive study and find salient factors that affect the adoption behavior. Our findings provide insights for app developers to improve their innovation and dissemination strategies. 
\end{itemize}

The rest of this paper is organized as follows. Section~\ref{sec:dataset} introduces the dataset we used in this study. Section~\ref{sec:diffusion} examines the adoption process of recurrent innovations in detail and proposes new categories of innovation adopters. 
Section~\ref{sec:regression} predicts the adoption decisions of recurrent innovations and investigates the factors that may influence the decision. 
Section~\ref{sec:regression-special} provides a further analysis of the new categories of adopters. 
Section~\ref{sec:discussion} discusses the limitations. Section~\ref{sec:related} introduces related work and possible research directions for interested readers. Section~\ref{sec:conclusion} concludes the paper.

\section{The Dataset}\label{sec:dataset}
To understand the adoption and diffusion process of repetitive innovations, we use a large-scale app usage dataset collected by \nickName, 
a leading third-party business intelligence service provider in China. This dataset records user behaviors in Android applications (i.e., apps) that utilize the \nickName 's SDK. 
We randomly sample 17,124,831 anonymous users who are active during the period from August 20, 2018 to June 30, 2019 and reconstruct their app usage sessions during the same time period. Each session is formulated as a 5-tuple, $<u, p, v, t_0, t_1>$, where $u$ and $p$ are anonymized identifiers for the user and the app, $v$ is the version of the app which the user is using, and $t_0$ and $t_1$ are the starting and ending timestamps of the session, respectively. Other than the beginning and the end of a session, we do not obtain any information about the actual activities or content of the session. 
We conduct a series of statistical tests to confirm that there is no significant difference between distributions the sample and the population. The sampled dataset includes 17,124,831 users, 217,285 apps, 713,935 app versions, and 17,013,616,656 sessions, covering a period from 20th Aug, 2018 to 30th June, 2019. 

\noindent\textbf{Ethical Consideration:} We took a series of steps to preserve the privacy of involved users in our dataset. First, all \textit{users} and \textit{apps} are anonymized by \nickName before made available to the authors. We know only the categories of the apps used by users, which means no users can be traced back through the data. Second, all data are kept within the \nickName 's private servers, which are protected by the company firewall. Additionally, the entire analysis is conducted on the servers of \nickName, which is strictly governed by \nickName 's administration. 

\section{The Adoption Process of Recurrent Innovations}
\label{sec:diffusion}

In a period of 10 months, the dataset records on average 3.28 versions per app, which provides a solid basis to study the diffusion of recurrent innovations. In this section, we provide a detailed examination of the adoption process of recurrent innovations, highlighting how it is both related to and different from the classic innovation diffusion theory.

For analogy, we treat each app as a unique \textit{technology} and a specific version of the app as a \textit{recurrent innovation} of that technology.  To illustrate the process with more concrete examples, we select one video app and one fitness app (referred to as "Video App" and "Fit App" thereafter), both are popular within their categories. 

\begin{figure}
    \centering
    \subfigure[Single Innovation \label{fig:single-innovation-sketch}]
    {\includegraphics[width=0.49\linewidth]{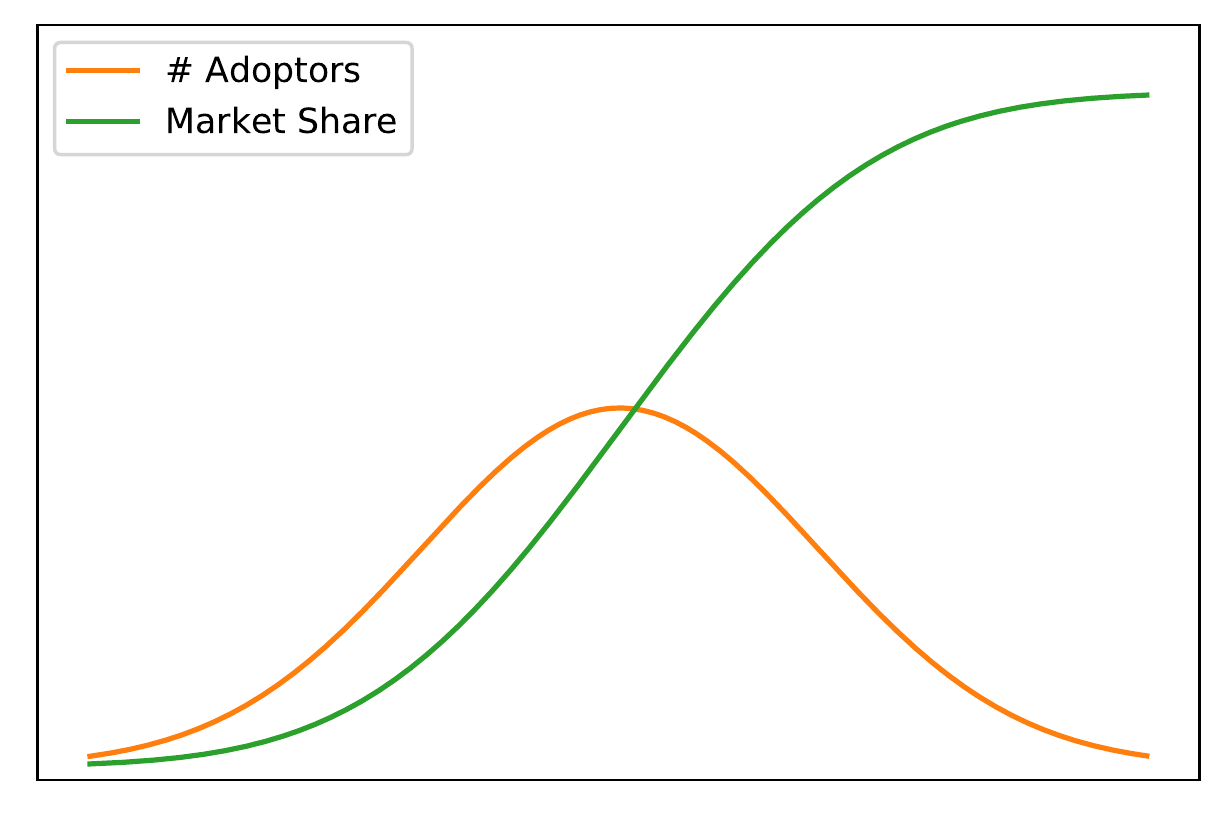}}
    \subfigure[Recurrent Innovation \label{fig:recurrent-innovation-sketch}]
    {\includegraphics[width=0.49\linewidth]{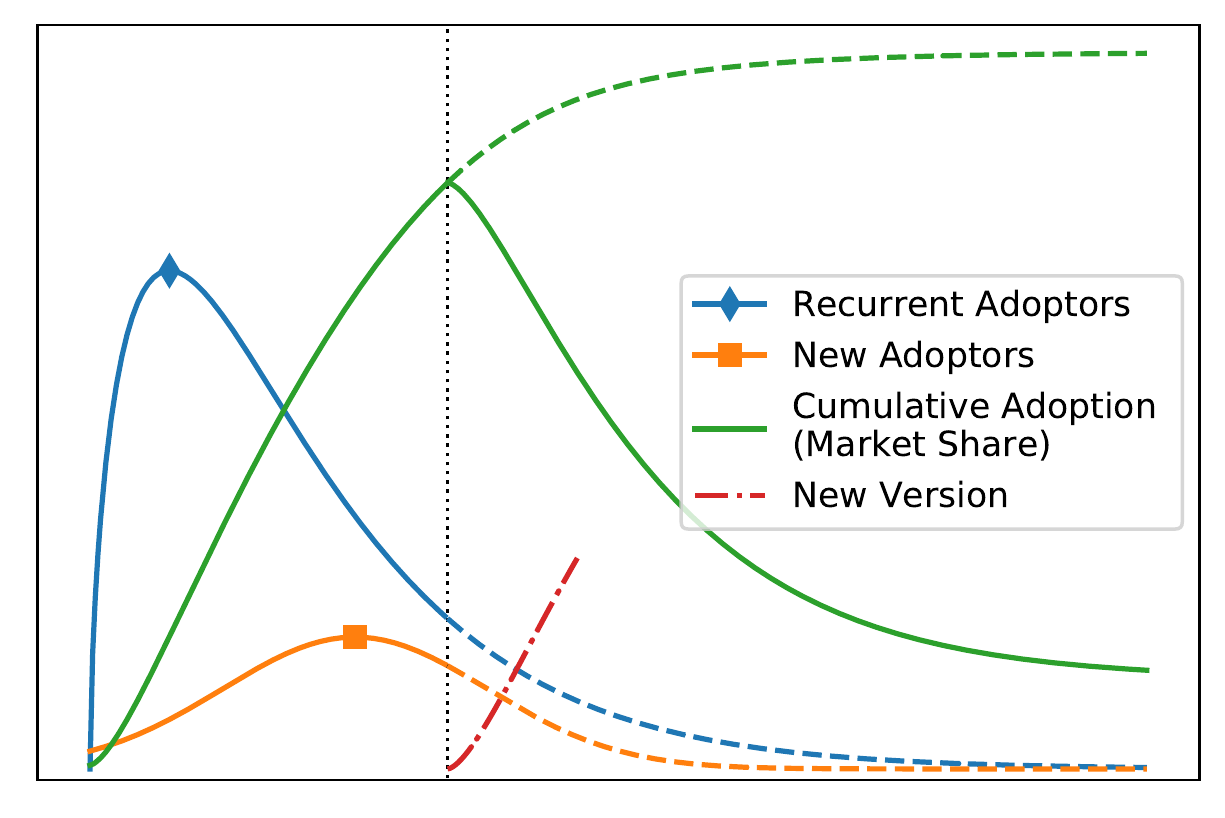}}
    \caption{Adoption curves for single innovation and recurrent innovation.}
    \label{fig:my_label}
\end{figure}

\subsection{Adoption Curves of Recurrent Innovations}
\label{subsec:time-of-adoption}

The classic innovation diffusion process in Rogers' theory is featured by the bell-shaped time-of-adoption curve (Figure~\ref{fig:single-innovation-sketch})\cite{rogers2010diffusion}. Users are categorized into \textit{innovators}, \textit{early adopters}, \textit{early majority}, \textit{late majority}, and \textit{laggards} according to their time of adopting a single innovation. In the context of fast evolving recurrent innovations, will the adoption curve still hold the same bell shape? Is there a difference between new adopters of the technology and those who have adopted an earlier version of the technology? Do the adopters of recurrent innovations fall into the same categories?  In this section, we start to examine the diffusion process of recurrent innovations through their time-of-adoption curves.

\begin{figure}
    \centering
    \subfigure[New Adopters\label{fig:add-line-new-video}]
    {\includegraphics[width=0.48\linewidth]{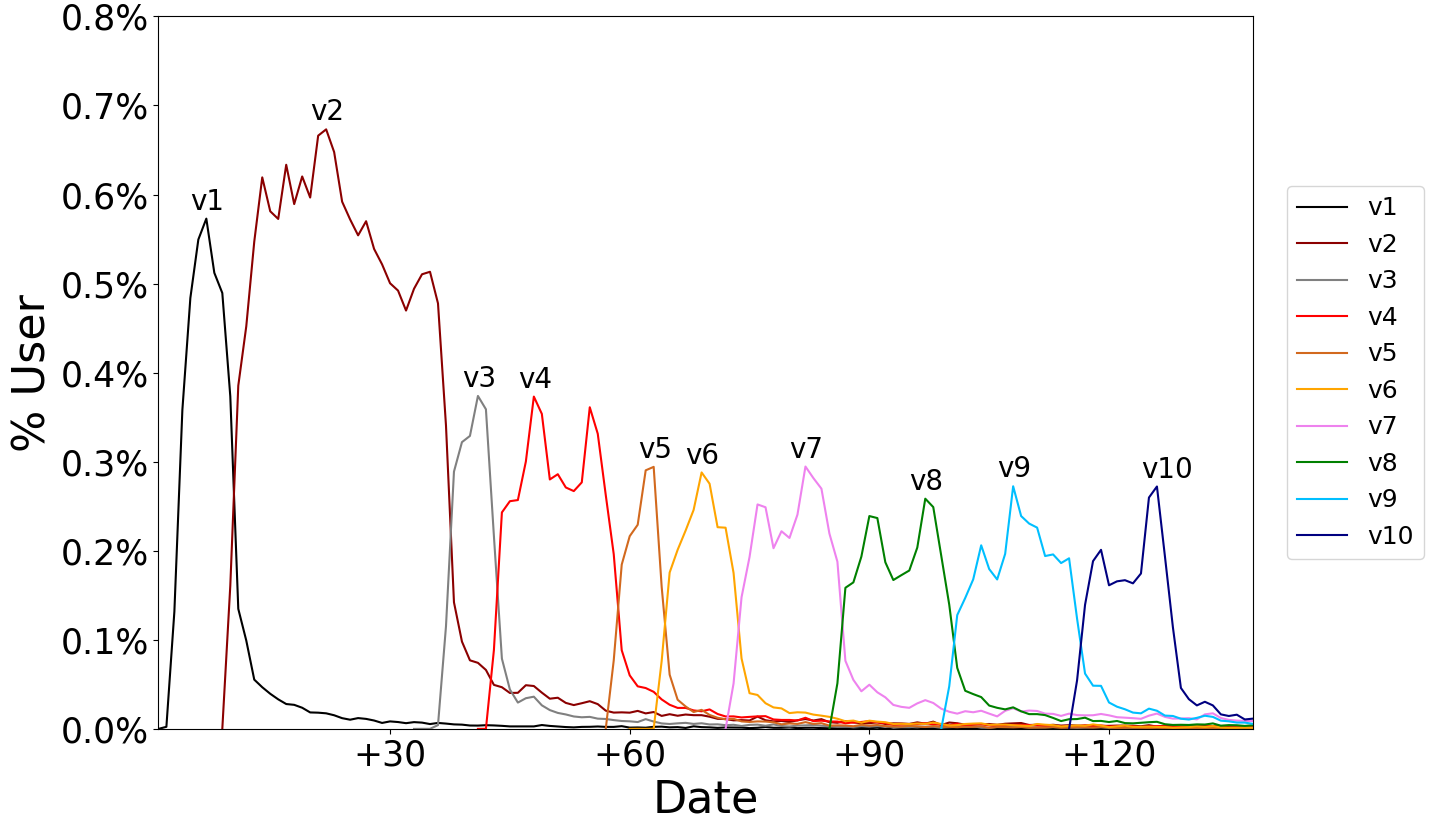}}
    \subfigure[Existing Adopters\label{fig:add-line-update-video}]
    {\includegraphics[width=0.48\linewidth]{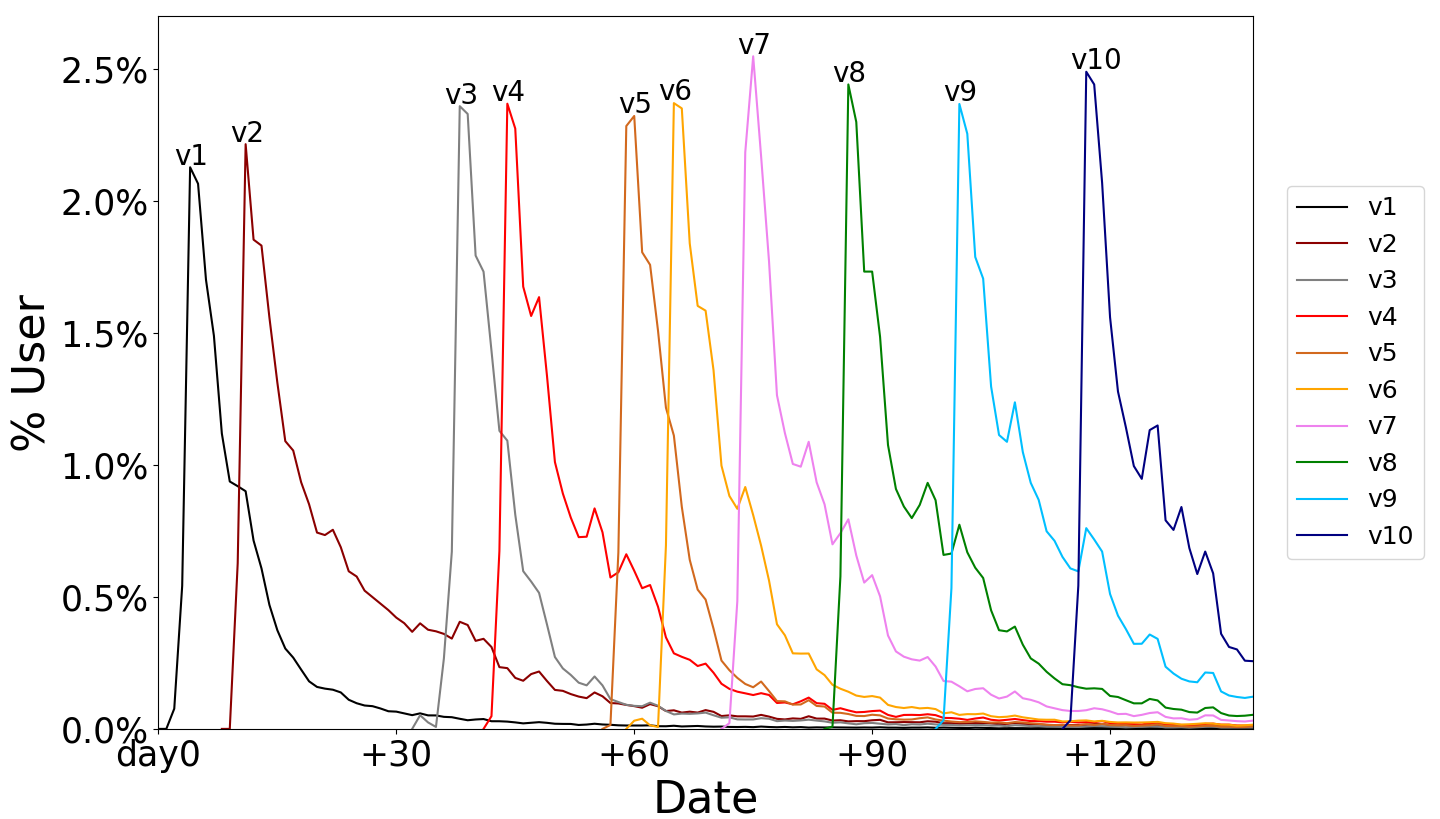}}
    \caption{Time-of-Adoption curve for each version of the Video App. The users are partitioned based on whether they have adopted an earlier version of the app (\emph{recurrent adopters}) or not (\emph{new adopters})}
    \label{fig:add-line-vedio}
\end{figure}

One thing we need to distinguish is that the same version of an app may be recognized as a recurrent innovation for some users, but as a brand-new innovation for other users. Indeed, for \emph{new adopters}, or users who haven't adopted any previous version of the app, a specific version of the app is not different from a single innovation, and we would expect a similar bell-shaped time-of-adoption curve for these users. In Figure~\ref{fig:add-line-new-video}, we plot the number of new adopters for each version of the Video App over a five-month period. We observe a bell-shaped curve for each version, despite some high-frequency jitters likely due to a weekly seasonality. Some new users get on board as soon as a new version just came out, some users wait to install the app later, while some users install an old version even after a newer version has already been released. The adoption curve of a specific version (a recurrent innovation) largely complies with the Rogers' curve for new adopters of the same app (technology).

Nevertheless, we observe the pattern that once a new version is started to diffuse, there is a sudden drop in the adoption curve of the precedent version. This indicates that even for new adopters, there is an influence between recurrent innovations. Not surprisingly, even a new adopter's decision depends on the timeliness of information, and the availability of a newer version would drive the attentions of users away from an older version. 

In comparison, the time-of-adoption curve for \emph{recurrent adopters}, or users who have adopted at least an earlier version of the app, exhibits a quite different pattern, as shown in Figure~\ref{fig:add-line-update-video}. Instead of a slow, graduate increase in the beginning, the curve rise sharply and exist a strong right-skewness. This indicates that we no longer observe the slow ramping-up of early adopters and the early majority. Instead, many recurrent adopters adopt responsively to the new version, and most recurrent adopters update to the new version within days of its release. Such adopters could be fans of the app and update it immediately when there is a new version, or they may have turned on the auto-update option for this app. Again, the adoption curve of a precedent version drops rapidly once a new version is released, although we can see some recurrent adopters update to a version even long after a newer version is released. 

The difference between recurrent adopters and new adopters indicate that there exist at least a new category of adopters. Comparing to early adopters in Rogers' theory, their decisions are even faster, in many cases even immediate. Instead of taking the reasonable staged processes of adoption (e.g., awareness, persuasion, decision, implementation, and continuation), these users make their adoption decisions in no time, almost blindly. We may call these users \textit{subscribers}. 

\subsection{Cumulative Adopters of an Iteration}
Following the Roger's theory, one intuitive derivation of the bell-shaped adoption curve is the S-shaped curve for cumulative adopters over time, which represents the market share of the innovation, as shown in Figure~\ref{fig:single-innovation-sketch}. That is, the total number of adopters increases slowly in the beginning (earlier adopters), faster afterwards (early/late majority), and slowly in the end (laggard). Yet with recurrent innovations, such S-shaped curve may never materialize. In fact, the cumulative adopters of a specific version cannot be calculated as simply as the integration of the time-of-adoption curve, since users may transit to newer versions and are no longer adopters of the current version.

\begin{figure}[htb]
    \centering
    \subfigure[Video App\label{fig:total-line-entertainment}]
    {\includegraphics[width=0.48\linewidth]{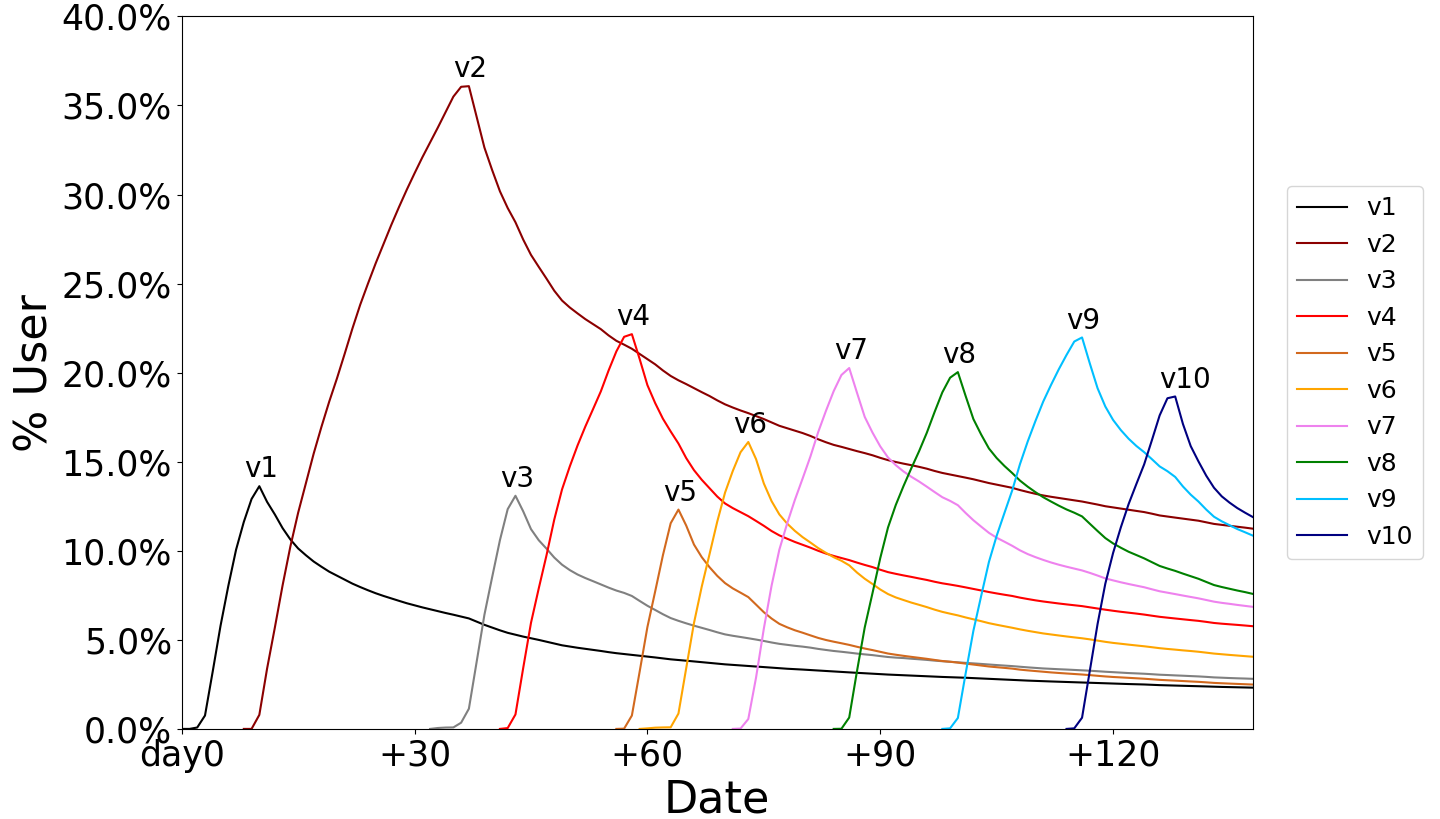}}
    \subfigure[Fit App\label{fig:total-line-fitness}]
    {\includegraphics[width=0.48\linewidth]{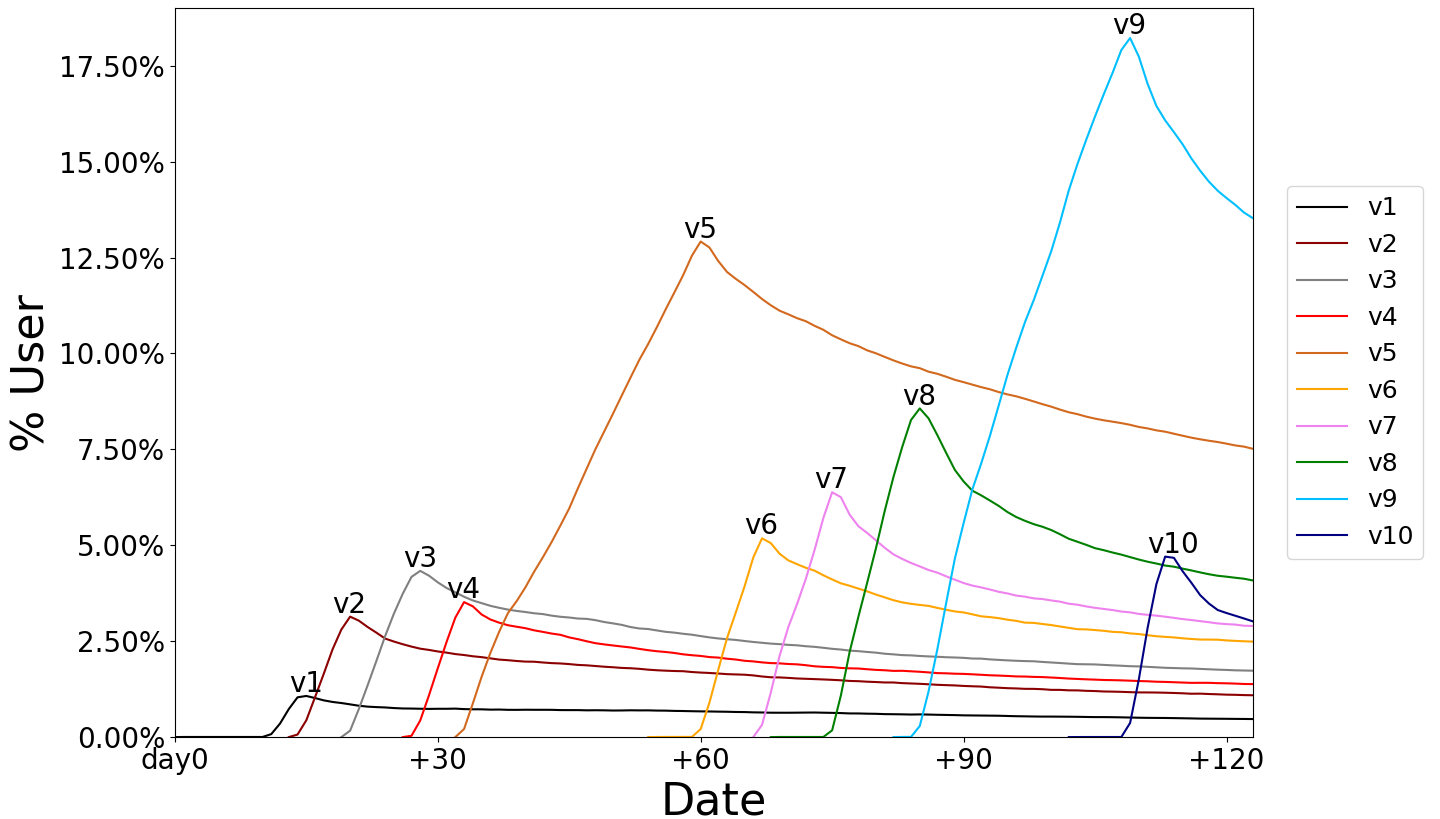}}
    \caption{Cumulative adopters of each version. The $y$-value of v-$i$ represents the number of users who have version $i$ of the app installed as of a given day, based on their most recent session of the app. 
    }
    \label{fig:total-line}
\end{figure}

The market share of iOS and Chrome in Figure~\ref{fig:market_share} sheds light on the cumulative adopters of recurrent innovations. The market share of a certain version rises upon release, and falls when a newer version is released. With the fine granularity of the app usage data, we can examine the adoption process on a daily level.  In Figure~\ref{fig:total-line}, we plot the number of users of different versions of the Video App and the Fit App over time, and we observe steady patterns of the adoption curve across different versions and different apps: 

Upon released, the number of adopters increases sharply, reaches its peak  when its successive version is released, and then declines, fast at first and slows down afterwards. The steep slope in both the increasing and decreasing phrase following the release of a new version is clearly driven by the recurrent adopters.

The curves eventually flat out but surprisingly always remain above zero, suggesting that a portion of adopters remain at the current version and never adopt newer versions of the same app.  This new categories of adopters are not covered by any theory of single innovations, and we may call them \textit{preservers} of an innovation. 

\subsection{New Types of Adoptors}

\begin{figure}[htb]
\centering
\includegraphics[width=0.8\linewidth]{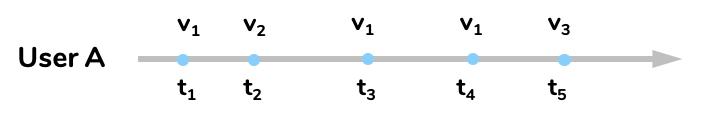}
\caption{An example of rolling back behavior. The user adopted an older version v1 after they adopted a newer version v2, making them a retro-adopter. }
\label{fig:roll-back}
\end{figure}

From the previous two subsections, we observe significant difference between the diffusion of recurrent innovations and that of single innovations. Specifically, we can identify several novel categories of users that only exist in the context of recurrent innovations, which are never covered in theories of single innovations: 

\begin{itemize}
    \item \textbf{Subscribers:} Recurrent adopters who promptly adopt a new innovation of the technology upon its release. We denote such users as \emph{the subscribers}.
    \item \textbf{Preservers:} Adopters of one innovation of the technology who do not adopt newer innovations. They keep an old version for a long time or even forever despite the availability of newer versions, whom we refer to as \emph{the preservers}.
    \item \textbf{Retro-adopters:} Although most users eventually adopt newer versions of the technology, some users decide to roll back to an earlier version after adopting a new version, as shown in Figure~\ref{fig:roll-back}. We refer to such users as \textit{retro-adopters}.
\end{itemize}

\textbf{It is worth mentioning that the three new categories, \emph{subscribers}, \emph{preservers}, and \emph{retro-adopters} have never covered by any existing theories}.  Indeed, the preservers and retro-adopters make sense only when there are multiple, recurrent innovations available.  Yet they represent a significant portion of users in our dataset. For example, about 57\% of versions gain more than 20\% subscribers, while about 0.5\% of versions gain more than 20\% retro-adopters. The proportion of preservers is even higher, i.e., about 90\% of versions gain more than 20\% preservers. Our analysis makes critical discoveries that call for an update of the diffusion of innovation theory in the context of recurrent innovations.  

\subsection{Uniting Recurrent and Single Innovations}

Despite the difference between recurrent innovations and single innovations, the two types of innovations are unified on a higher level. In a longer time frame, if we view an app as an innovation, then the adoption of this innovation is made up of the adoptions of each of its versions. In other words, the diffusion of a technology overtime can be decomposed into the the diffusions of all its recurrent innovations - except that they are interweaving and interacting with each other.  Our analysis provides a lens for understanding the rises and falls of these recurrent innovations and their interactions at a finer granularity.  

\begin{figure}[htb]
    \centering
    \begin{center}
        \subfigure[Fitness App\label{fig:total-stack-fitness}]
        {\includegraphics[width=0.48\linewidth]{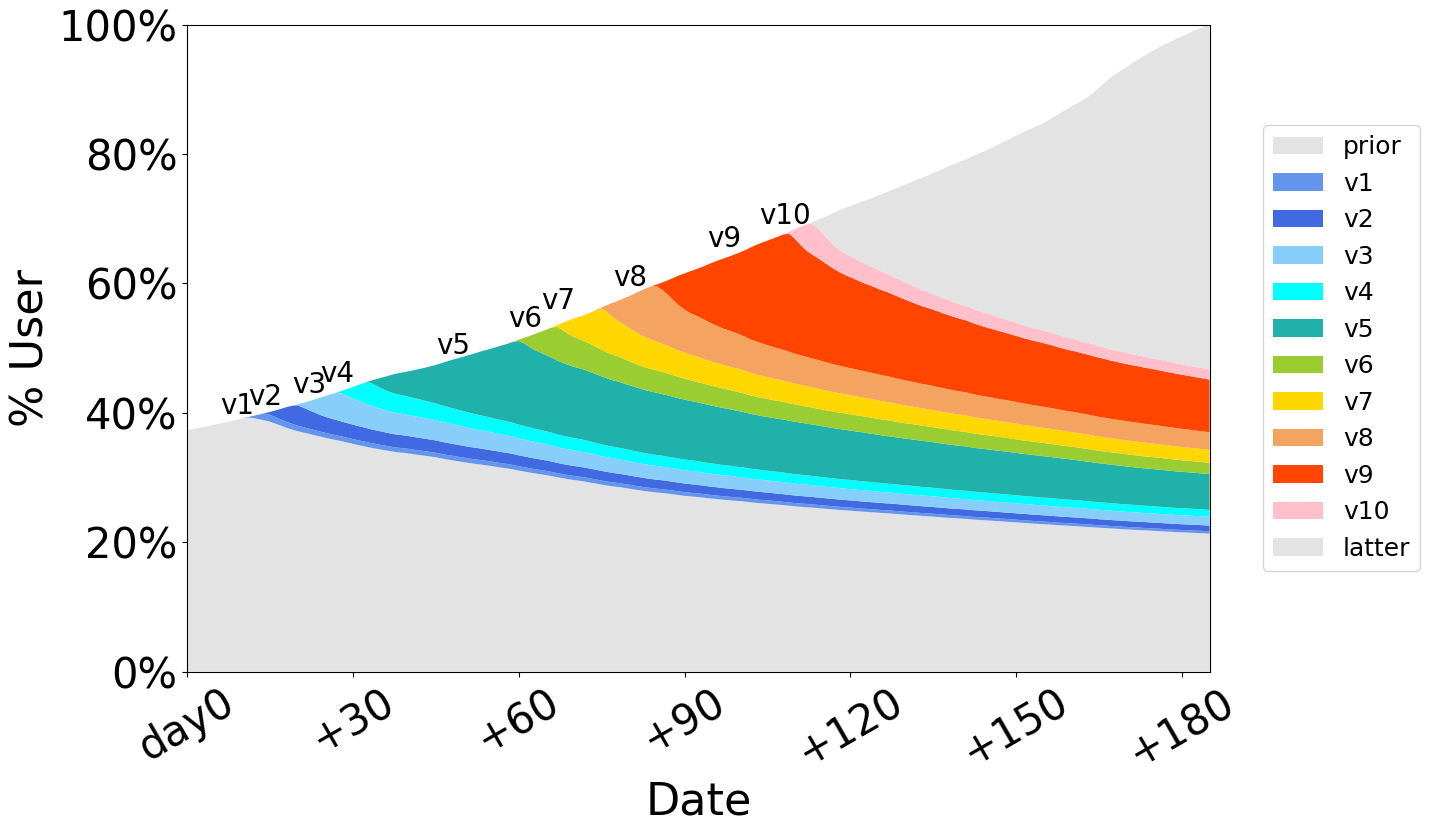}}
        \subfigure[Video App\label{fig:total-stack-video}]
        {\includegraphics[width=0.48\linewidth]{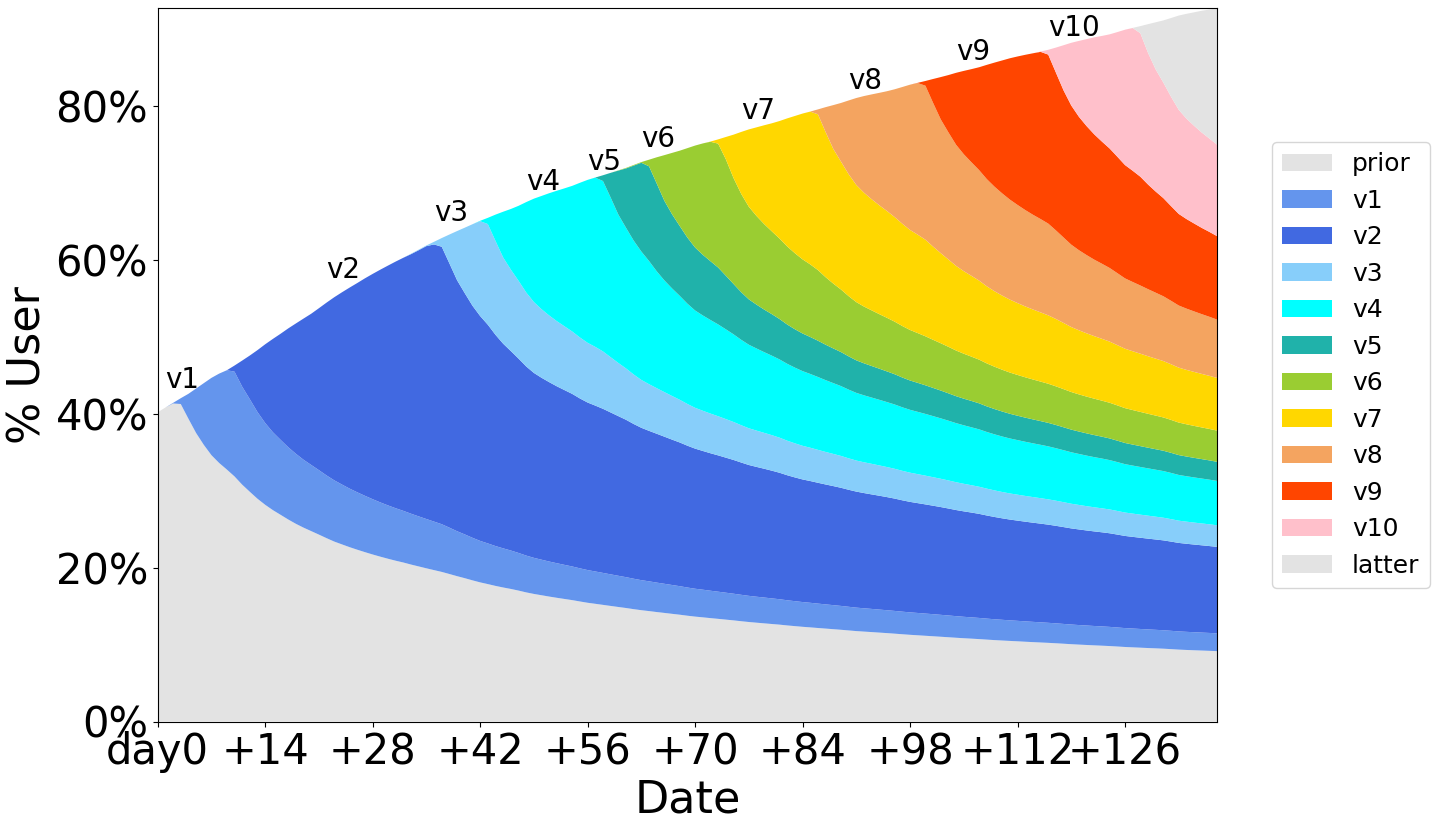}}
        \caption{Cumulative adopters over multiple versions. Recurrent versions contribute to the adoption curve of the app as a whole, but there exist considerable variations among versions.  }\label{fig:total-stack}
    \end{center}
\end{figure}

Indeed, Figure~\ref{fig:total-stack} demonstrates the cumulative adopters of an app with the adopters of every version stacked on top of each other. At any time point, the height of each layer represents the market share of the corresponding version, and the height of all layers represents the users who have adopted the app. The stacked plot confirms the dual roles of a recurrent innovation. On one hand, it transits existing adopters to the latest version (or occasionally to an earlier version), and on the other hand, it draws new adopters and contributes to the increasing adoption of the technology as a whole. The envelope curve of the stacked layers indicates the cumulative adoption of the app in the observed population. However, such a curve does not necessarily follow the the S-shape of that of a single innovation, since individual innovations of the technology (individual versions of an app) may be differently appealing, or they may appeal to different segments of the population (because of the differences in features, bugs, or operating strategies), and this difference opens up new opportunities for innovators, disseminators, and even competitors.  

To summarize, we observe novel and intriguing patterns in the adoption curves of recurrent innovations that the classical innovation adoption theory cannot explain. Specifically, we suggest three new categories of adopters to be added into the diffusion of innovation theories, which are \textit{the subscribers}, \textit{the preservers}, and \textit{the retro-adopters}. While the adoption curves of recurrent innovations are more or less alike, there exist variations among different versions of the same app, which implies that the adoption decisions vary per the individual user and per the individual innovation. In the next section, we explore factors that may predict the individual decisions of adopting a particular version of the recurrent innovation.

\section{Individual Adoption Decisions}
\label{sec:regression}

The analysis in Section~\ref{sec:diffusion} reveals the difference of adoption behavior in different types of users and different versions of an app.  To further understand the variance in the individual decisions of adopting a recurrent innovation, we are interested in a predictive analysis. Is a user's behavior of adopting a recurrent innovation predictable? If yes, what kind of factors could explain the decision of adoption? To answer such questions, we design a prediction task to analyze the users' adoption decisions, that is, whether and how soon a user will adopt a specific version. 

Given the time span of the data collection, we set up the prediction tasks as follows: for each app that has released a new version in May 2019, we sample its users who have actively used the app within 30 days prior to the release and track whether they will use this new version within 30 days (implying they have installed it), and if yes, when they use it for the first time.

We first introduce the list of features that comprehensively represent the characteristics of the app, the user, and interactions between the user and prior versions of the app. 

\subsection{Feature Extraction and Selection}
\begin{table*}[]
\caption{Features}
\begin{tabular}{l|l}
\hline
Dimension & Features \\\hline
\begin{tabular}[c]{@{}l@{}}D1: properties of the\\  target app\end{tabular}           & \begin{tabular}[c]{@{}l@{}}\#used days, avg. daily users, \#monthly users, avg. daily co-used apps, \#versions a month, \\ \#versions a quarter, avg. release interval, time of release\end{tabular} \\\hline
\begin{tabular}[c]{@{}l@{}}D2: the user's \\ interaction with \\ the target app\end{tabular} & \begin{tabular}[c]{@{}l@{}}prop. days using the app in the last week/two weeks/one month, \#times using the app \\ in the last week, time spent in the app in the last week;\\ \#days to adopt the last/second last/fourth last/eighth last version, \\ freq. adopting the last eight versions, std. days to adopt the last eight versions\end{tabular} \\\hline
\begin{tabular}[c]{@{}l@{}}D3: characteristics \\ of the user\end{tabular}            & \begin{tabular}[c]{@{}l@{}}\#used apps, \#active days, \#times using apps, time spent in all apps, avg. days using other \\ apps, avg. times using other apps, avg. time spent in other apps;\\ \#versions adopted, avg. days to adopt a version by app/version\end{tabular}   \\\hline
\end{tabular}
\label{tab:features}
\end{table*}

With the fine granularity in the dataset, we can extract three sets of features, as shown in Table~\ref{tab:features}. All features are calculated using data within 30 days prior to the release of the target version unless otherwise noted.

The first set of features represent properties of the innovation, that is, the app (F1 in Table~\ref{tab:features}). These include its popularity (\textit{the (average) daily active users}, and \textit{monthly active users}. 
We also extract the average number of apps that the users of this app interact with on the same day (\textit{the number of co-used apps}). 
We also characterize the updating history of the app, that is, the number of versions released in the recent one month and three months (\textit{number of versions}), and average interval of the last three releases (\textit{the intervals of the last three versions}). Finally, since the in-app usage differs by time~\cite{liu2017understanding,DBLP:journals/tweb/MaHGZMHL20}, the timing of releases might also affect the adoption behavior, so we encode the day-of-the-week and hour-of-the-day of the release as one-hot features.

The second set of features describe the \emph{adopters}. 
In specific, we record how heavily a user uses their phone, by measuring \textit{the number of apps} they used, \textit{the number of days} that they use any apps, \textit{the frequency of launching apps}, and \textit{the total time spent in apps}. 
We further measure their per-app usage by measuring the number of days, length, and launching frequency of other apps except the one in focus.
In addition, we measure the user's prior interaction with recurrent innovation by measuring \textit{the number of versions they adopted} and \textit{the average days to adopt a version}.

Given the recurrent nature of the release, the last set of feature characterize the users' interaction with the app prior to the release, including both how they use the app and how they adopt prior versions of the app. 
The former includes \textit{the proportion of days using the app} in the recent one week, two weeks, and one month, respectively; \textit{the frequency of launching the app} and \textit{the total time spent in the app} in the recent week. The latter includes \textit{the adoption interval} between the release and the adoption of the 1st, 2nd, 4th, and 8th most recent update. In addition, we calculate \textit{the likelihood of adopting the last eight versions} and \textit{the standard deviation of the adoption interval}.

\subsection{Experiment Setups}
\subsubsection{Data Processing}
We take a few steps in executing the experiment to ensure the internal validity. Readers may refer to the Appendix for the detailed pre-processing. Eventually, we are able to select 1,651 app versions, and 1,063,244 observations for the ``adopt-or-not'' classification, among which 222,432 observations are positive and thus used in the ``time-to-adopt'' regression. For both tasks, we randomly split the dataset into 80\% training and 20\% test set, and apply 5-fold cross-validation on the training set for tuning hyper-parameters.

\begin{table*}[]
\centering
\caption{The dataset size of each task}
\label{table:dataset-size}
\begin{tabular}{|c|c|c|c|c|c|c|}
\hline
\multicolumn{1}{|l|}{} & \multicolumn{5}{c|}{Classification} & \multicolumn{1}{l|}{Regression} \\ \hline
Task & Adoption & The Conservatives & \multicolumn{2}{c|}{The Innovation Followers} & The Roll-Back Adopters & Adoption \\ \hline
Size & 1,063,244 & 202,266 & 222,432 & 222,432 & 80,000 & 222,432 \\ \hline
\end{tabular}
\end{table*}

\subsubsection{Feature Selection}

In practice, some features are likely to be highly correlated with ones another, leading to multicollinearity issues that deteriorate the prediction models the interpretation of the linear models. Following the standard practice, we calculate the pair-wise Pearson coefficient between all pairs of features (except the one-hot features) on the training sets and carefully select features so that the correlation coefficients between any of them are not higher than 0.7~\cite{dormann2013collinearity}. We repeat the process for both the classification and the regression task. Eventually, we have 17 features for both tasks.

\begin{figure}[htb]
    \centering
    \begin{center}
        \subfigure[Importance\label{fig:classification-importance}]
        {\includegraphics[width=0.48\linewidth]{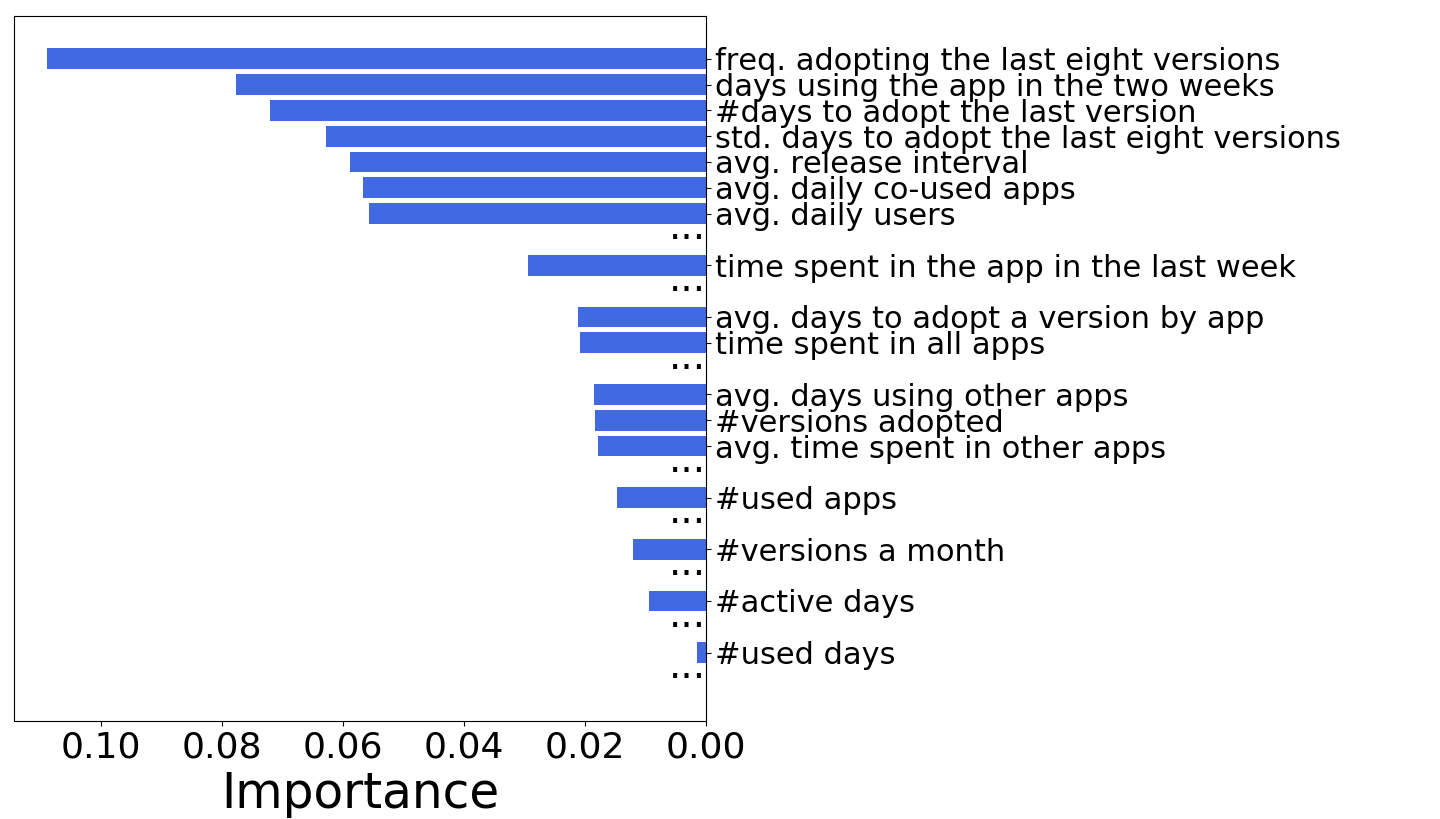}}
        \subfigure[Logistic Regression Coefficient\label{fig:classification-coefficient}]
        {\includegraphics[width=0.48\linewidth]{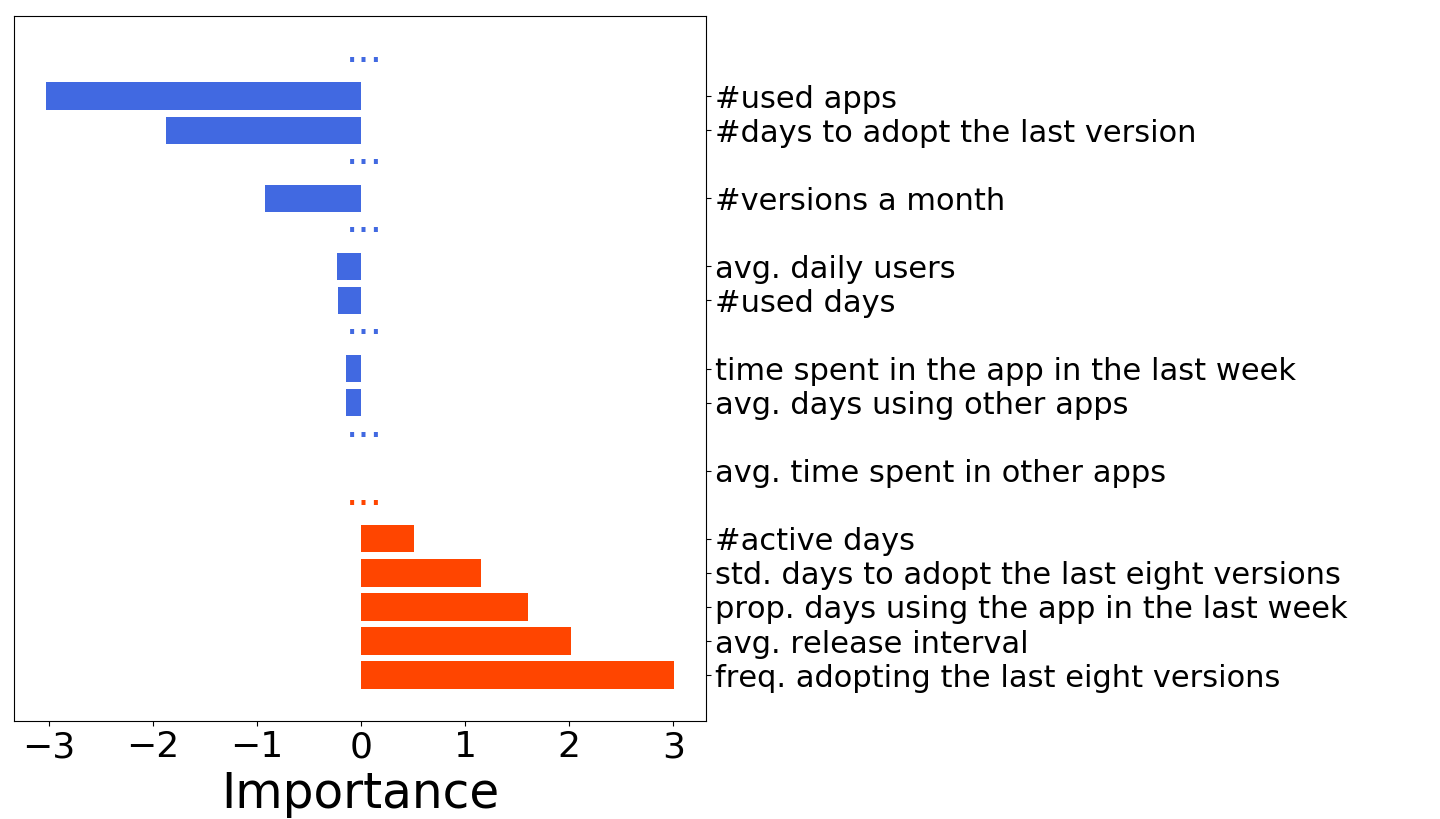}}
        \caption{The Importance and Coefficient of Features in Classification.}\label{fig:classification-coef-im}
    \end{center}
\end{figure}

\begin{figure}[htb]
    \centering
    \begin{center}
        \subfigure[Importance\label{fig:regression-importance}]
        {\includegraphics[width=0.48\linewidth]{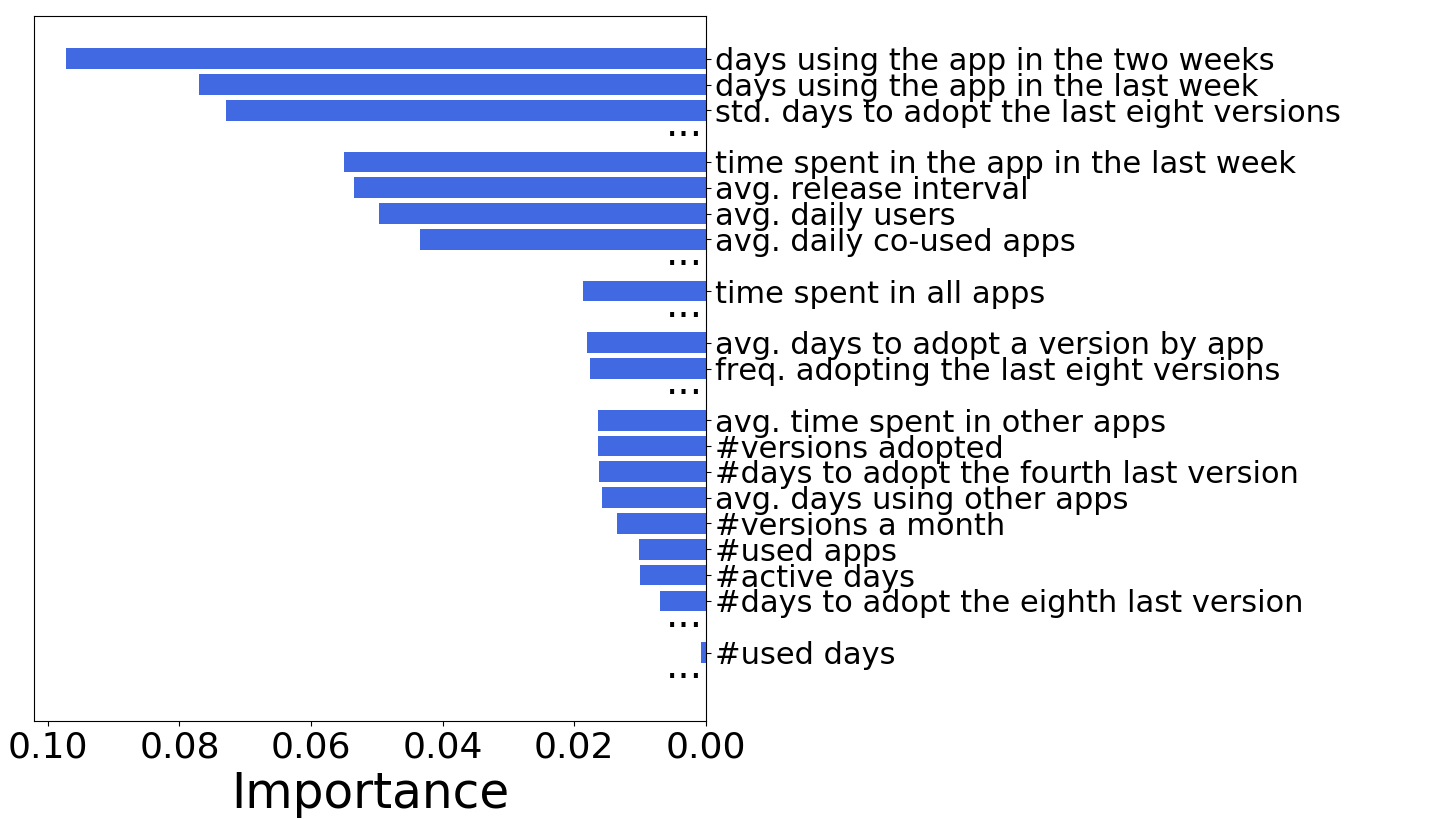}}
        \subfigure[Linear Regression Coefficient\label{fig:regression-coefficient}]
        {\includegraphics[width=0.48\linewidth]{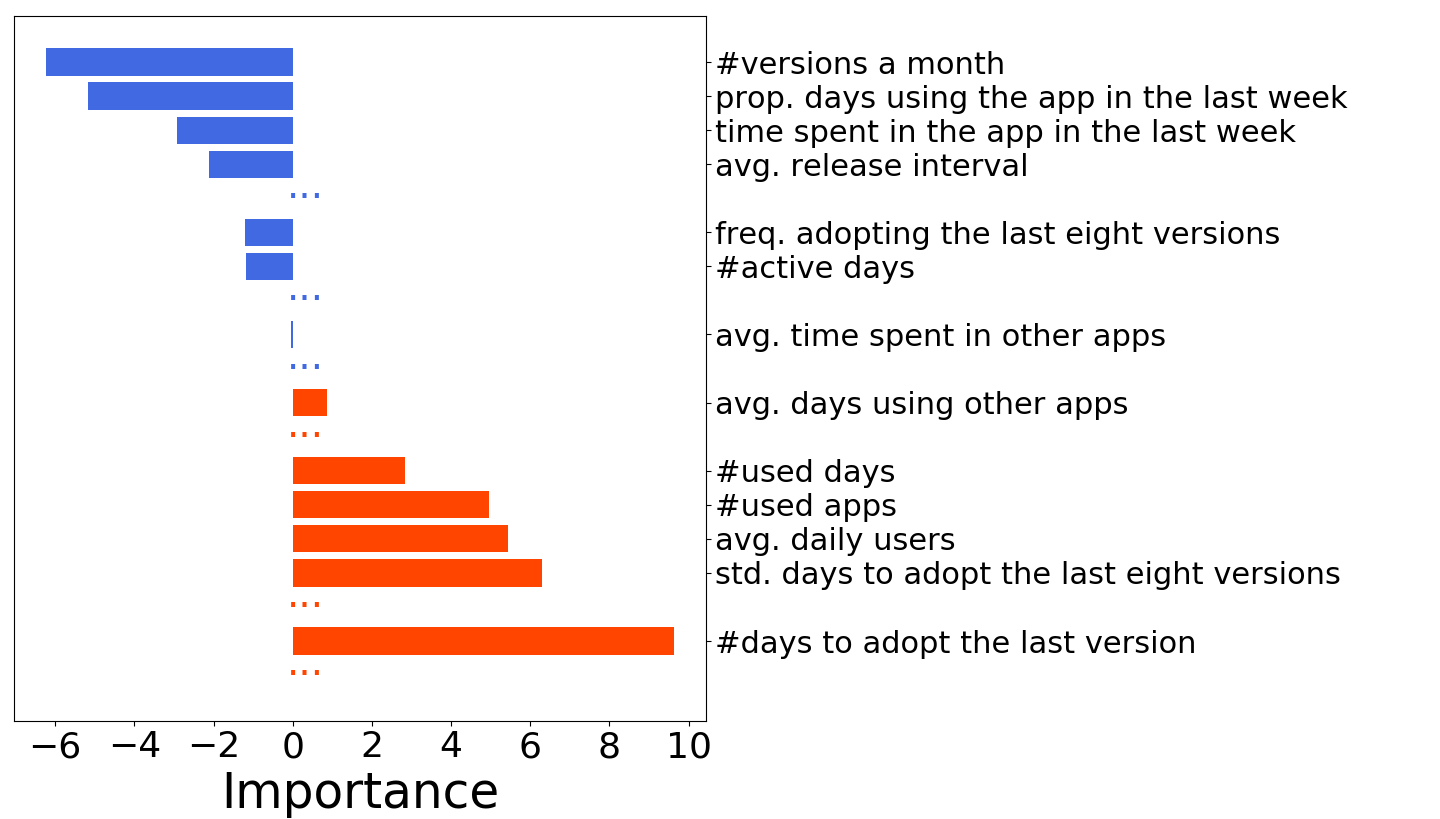}}
        \caption{The Importance and Coefficient of Features in Regression. Negative correlation to \textit{\#days to adopt} indicates sooner adoption.}\label{fig:regression-coef-im}
    \end{center}
\end{figure}

\subsubsection{Model Selection}\label{subsubsec:setups}

For the classification task, we select two  models, i.e., Gradient Boosting Decision Tree (a.k.a., GBDT) and Logistic Regression. 
For the regression task, we select four different models, i.e., GBRT, OLS regression, Ridge regression, and Lasso. We use LightGBM\footnote{\url{https://lightgbm.readthedocs.io/en/latest/}} for GBDT/GBRT and  statsmodels\footnote{\url{https://www.statsmodels.org/stable/index.html}} for other models.

\subsubsection{Evaluation Metrics}

For the classification models, we select Area Under Curve (a.k.a., AUC) score and accuracy to evaluate the model performance. A larger AUC or a larger accuracy indicates better classification results. We select Root Mean Squared Error (a.k.a., RMSE) score and R-Square score as the metrics for the regression models. A smaller RMSE score and a larger R-Square score indicate a better model in the regression tasks. We leverage the majority guess as the baseline. 

\subsection{Results}
We next report the expreimental results. 

\subsubsection{Model Performance}

\begin{table}[]
\caption{Results of the Classification of Adoption Tasks}
\label{table:clf-adoption-result}
\begin{tabular}{|c|r|r|}
\hline
Model & \multicolumn{1}{c|}{AUC} & \multicolumn{1}{c|}{Accuracy} \\ \hline
GBDT & 0.8813 & 0.8488 \\ \hline
LR & 0.7542 & 0.7970 \\ \hline
Baseline & 0.5000 & 0.7682 \\ \hline
\end{tabular}
\end{table}
\begin{table}[]
\centering
\caption{Results of the Regression of Adoption Tasks}
\label{table:reg-result}
\begin{tabular}{|c|c|c|}
\hline
Model & RMSE & R Square \\ \hline
GBDT & 6.1936 & 0.2685 \\ \hline
Linear Regression & 6.8329 & 0.1097 \\ \hline
Ridge & 6.8330 & 0.1097 \\ \hline
Lasso & 6.8662 & 0.1010 \\ \hline
\end{tabular}
\end{table}


As presented in Table~\ref{table:clf-adoption-result}, the GBDT and LR models both beat the baseline model in terms of  AUC and accuracy. The GBDT model performs the best with the AUC of 0.8813 and the accuracy of 0.8488, respectively. Such a result indicates the power of the  proposed features in distinguishing adopters and non-adopters of a new version.

Table~\ref{table:reg-result} presents the results of regression task that predicts how long it takes an adopter to update to a version. In all four models, the GBDT model achieves the best performance with an R-Square of 0.2685 and an RMSE of 6.1936, respectively. Such a result implies that the GBDT model is well adequate to our proposed features. The RMSE  predicting the number of days after the release of a version is around 6, indicating the period is within a week.

After demonstrating the predictive powers of the selected features, we further explore the significance of features in the prediction task and further investigate their relation with adoption.

\subsubsection{Feature Interpretation} 
We rely on the feature importance scores that are  generated by the GBDT models to select. we report the coefficients of features from the LR models  in Figure~\ref{fig:classification-coef-im} and~\ref{fig:regression-coef-im}, respectively.

It can be observed that quite a few features are both important in the two tasks. It is reasonable as the two tasks reflect the attitude and decision of adopting a version. In addition, the most important features in the two tasks are those who describe how users interact with the innovation (i.e., apps), with slight difference in ranking. As for the ``adopt-or-not'' classification, the most important feature, i.e., the frequency of adopting the last eight versions, indicates one's tendency to adopt versions of this app, or innovativeness. The coefficient with a positive sign further indicates that users adopting more history versions of an app also tend to adopt a new version of it (evidence of \textit{recurrent adopters}). Meanwhile, the most important feature in the ``time-to-adoption'' regression is the number of days one used the app in the last two weeks, followed by the number in the last one week. As is reported in Figure~\ref{fig:regression-coefficient}, such features are negatively related to the time to adopt, indicating that if is to adopt, a user that use the app more frequently adopt the new version sooner (evidence of \textit{subscribers}). 

Following are features from the first dimension that describes the properties of the target app, among which the average release interval is the most important for both the tasks. A larger release interval, which means a less frequent release pattern, is related with a higher tendency and shorter interval of the users to adopt the version. This finding provides direct insights for mobile app developers in their design of release plans.

Features to characterize the users, including the number of used apps, the number of active days, the time spent in all apps, etc., also show their predictive power. Users using more apps are less likely to adopt a new version and if they do, it would take more time for them to adopt it (evidence of \textit{preservers}, and that preservers might preserve multiple apps). More active days of using any of the apps represents the frequency that one uses the mobile phone, which is positively related with the adoption decision and negatively related with the days to adopt (a faster adoption). 

These findings reveal that in addition to inherent characteristics of the innovation and the adopters, the patterns in the previous versions of a recurrent innovation including how they were released, how users adopted them, and how adopters interacted with them, can affect users' decision of adoption. 

\section{Understanding the Special Adopters}\label{sec:regression-special}

As is found in Section~\ref{sec:diffusion}, there exist three different groups of adopters (i.e., the subscribers, the preservers, and the retro-adopters) of recurrent innovations. The characteristics of such adopters, i.e., adopt or reject an innovation in a different way, provide a chance to mitigate the pro-innovation bias~\cite{rogers2010diffusion}, which means that all innovation are assumed positive and should be adopted, and the individual-blame bias, which means a tendency for diffusion research to side with the innovation creators while ignoring the audience.  The subscribers who adopt the recurrent innovation in a short time show quite positive attitudes towards the innovation. The preservers who stay with the innovation show their loyalty to it or negative attitudes towards the subsequent innovations since they refuse to adopt them. As for the retro-adopters who roll back to previous versions, they show clear negative attitudes towards the new innovation. In other words, rather than other adopters, behaviors of adopters in these three categories shed lights on the understanding of polarized opinions towards innovations from the crowd. 

The results in Section~\ref{sec:diffusion} shows some evidence of the existence and difference of subscribers and preservers. The setup of the tasks does not allow us to investigate retro-adopters. To better understand the behaviors of the special categories of adopters, and hopefully the reasons behind their adoption behaviors, we conduct a correlation analysis between the features we proposed in Section~\ref{sec:regression} and the three types of adopters. 
Note that we use the 1,651 versions and corresponding user samples as in Section~\ref{sec:regression} to analyze the subscribers and the preservers, while the total 4,995 versions to analyze the retro-adopters. Interested readers can refer to Appendix~\ref{appendix:types} for details of detecting each type of adopters.

\begin{figure*}[htb]
    \centering
    \begin{center}
        \subfigure[All users\label{fig:all-users-last-interval}]
        {\includegraphics[width=0.24\linewidth]{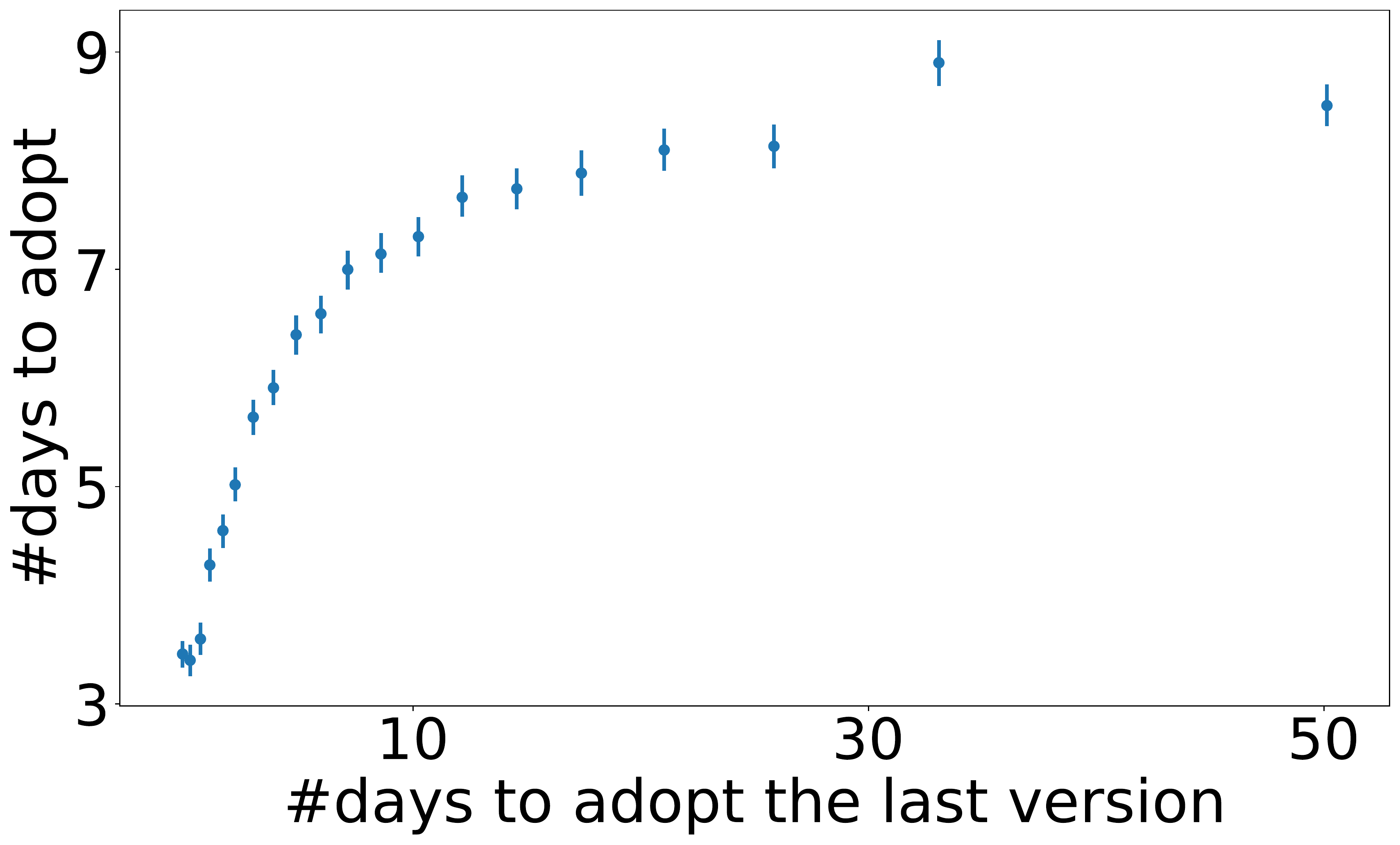}}
        \subfigure[Subscribers\label{fig:subscribers-last-interval}]
        {\includegraphics[width=0.24\linewidth]{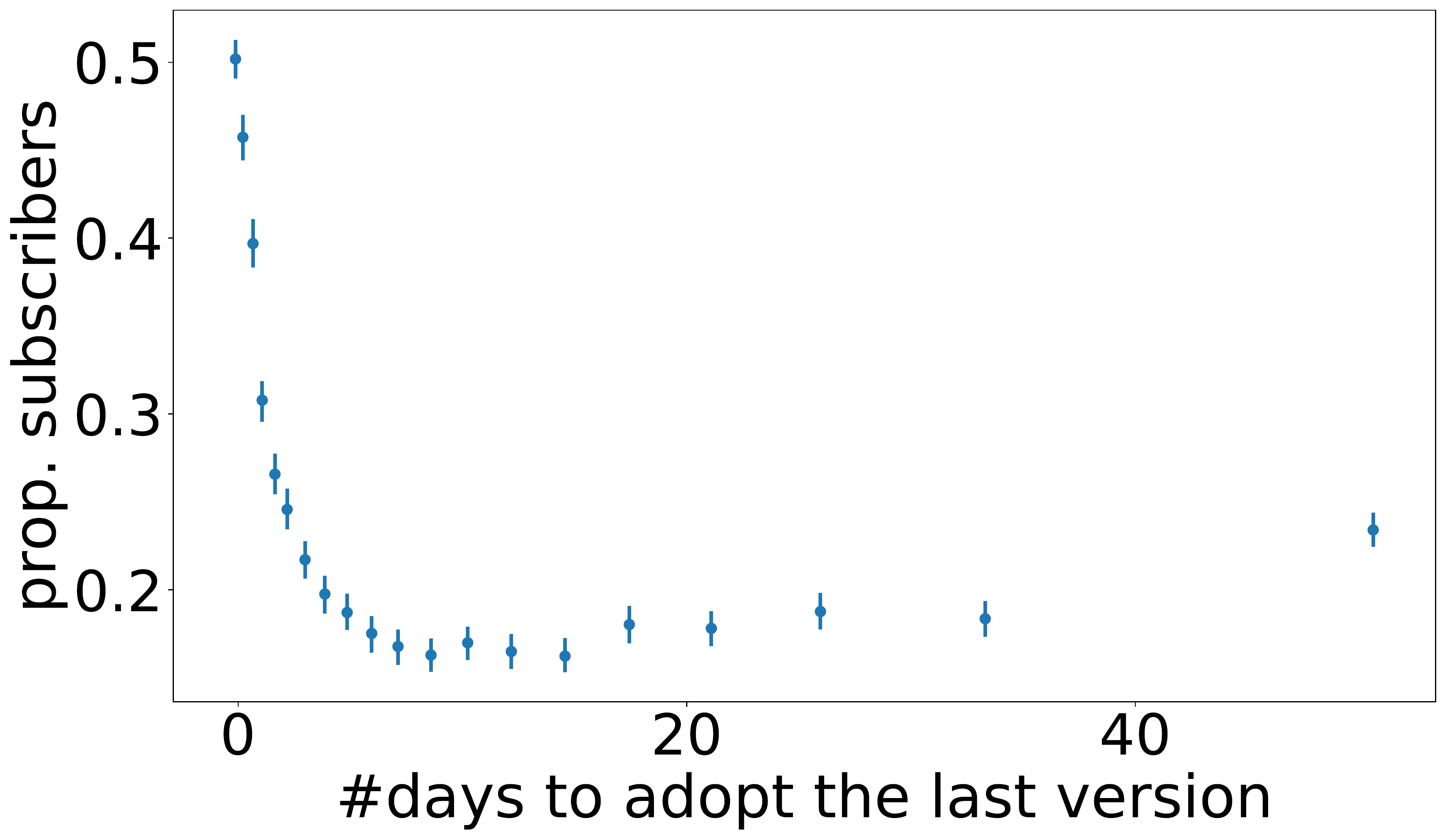}}
        \subfigure[Preservers\label{fig:dwellers-last-interval}]
        {\includegraphics[width=0.24\linewidth]{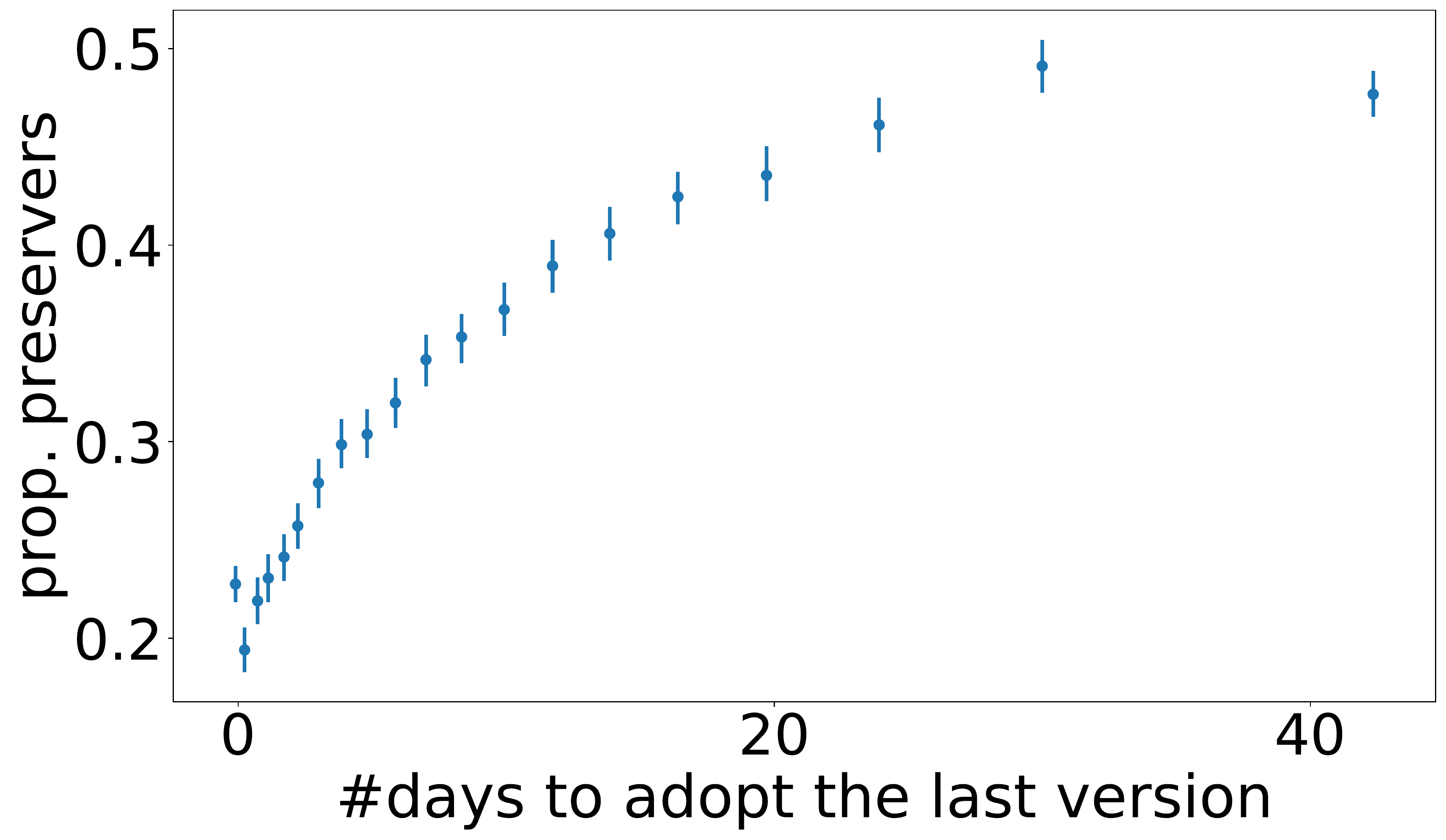}}
        \subfigure[Retro-adopters\label{fig:retro-adopters-last-interval}]
        {\includegraphics[width=0.24\linewidth]{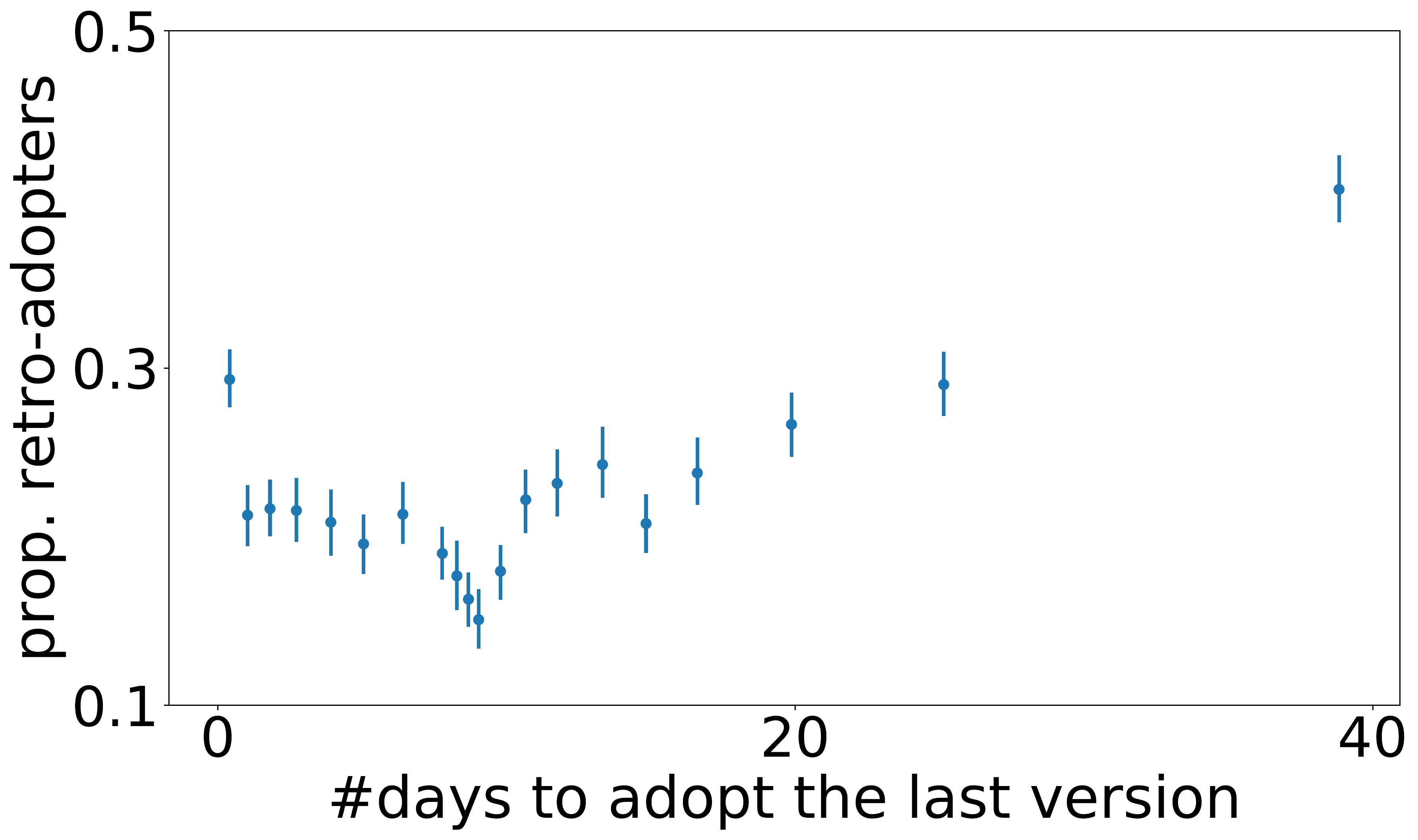}}
        \caption{Correlating adoption with previous adoption.}\label{fig:last-interval}
    \end{center}
\end{figure*}

\begin{figure*}[htb]
    \centering
    \begin{center}
        \subfigure[All users\label{fig:all-users-last-week-duration}]
        {\includegraphics[width=0.24\linewidth]{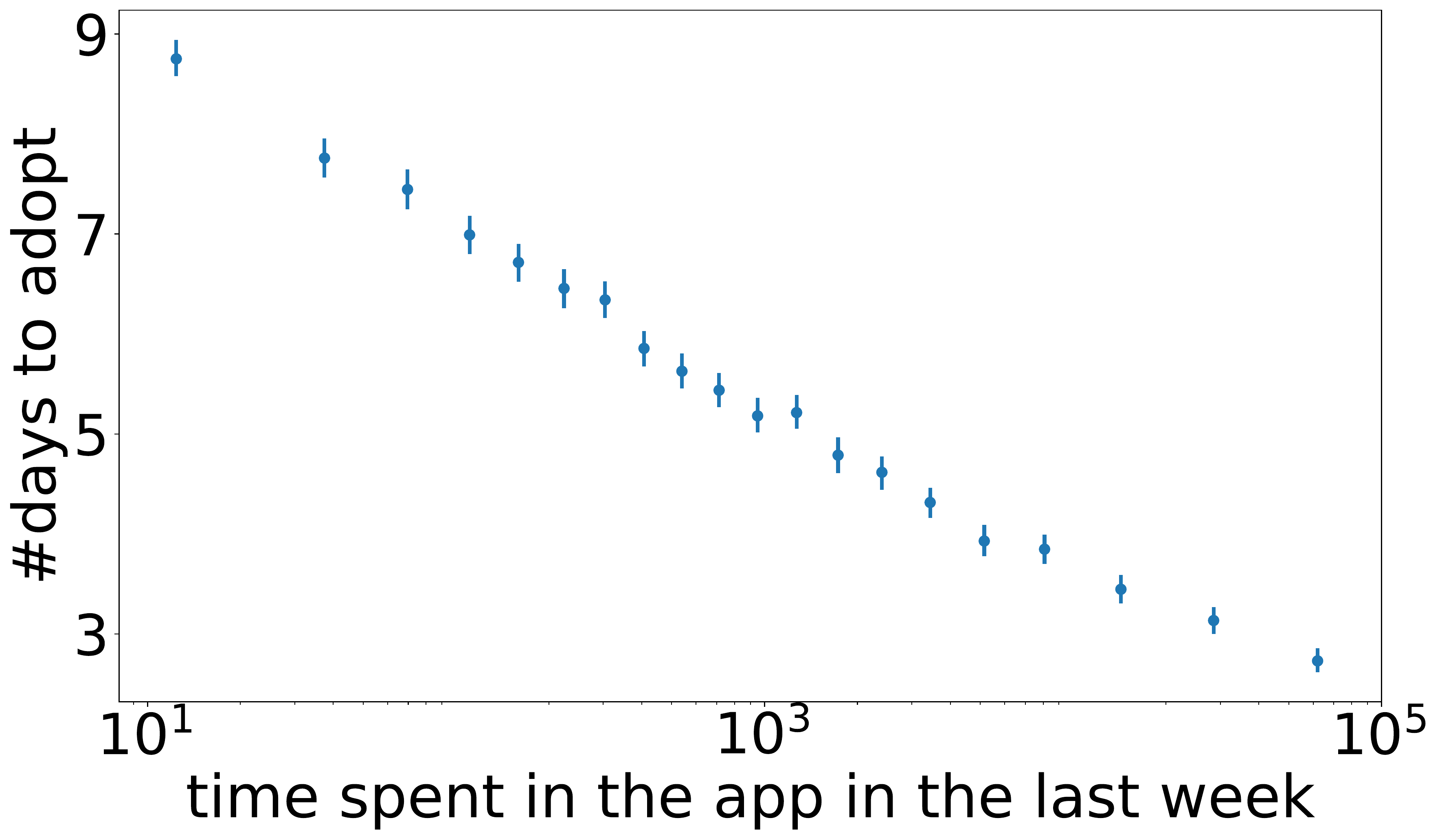}}
        \subfigure[Subscribers\label{fig:subseribers-last-week-duration}]
        {\includegraphics[width=0.24\linewidth]{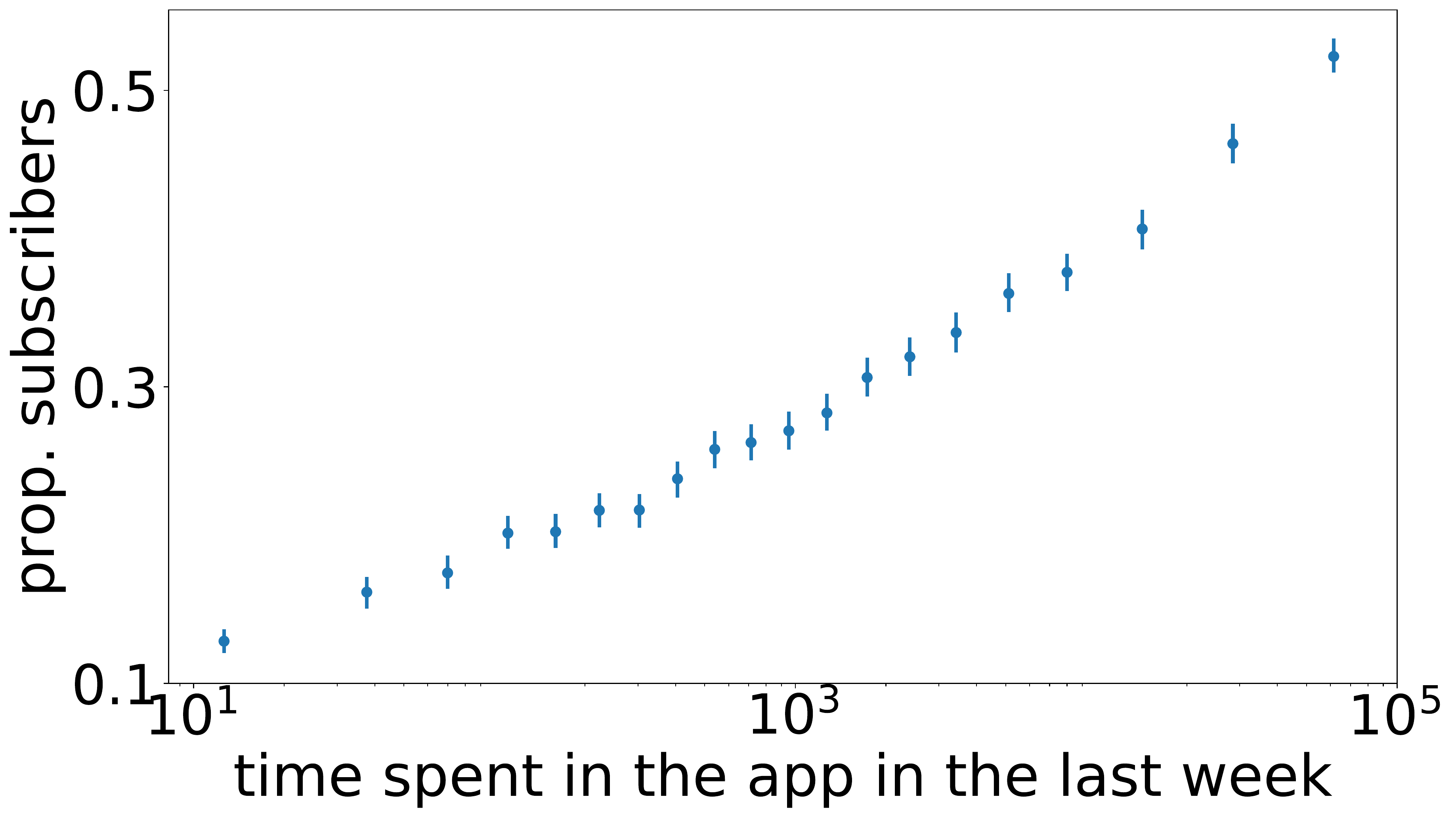}}
        \subfigure[Preservers\label{fig:dwellers-last-week-duration}]
        {\includegraphics[width=0.24\linewidth]{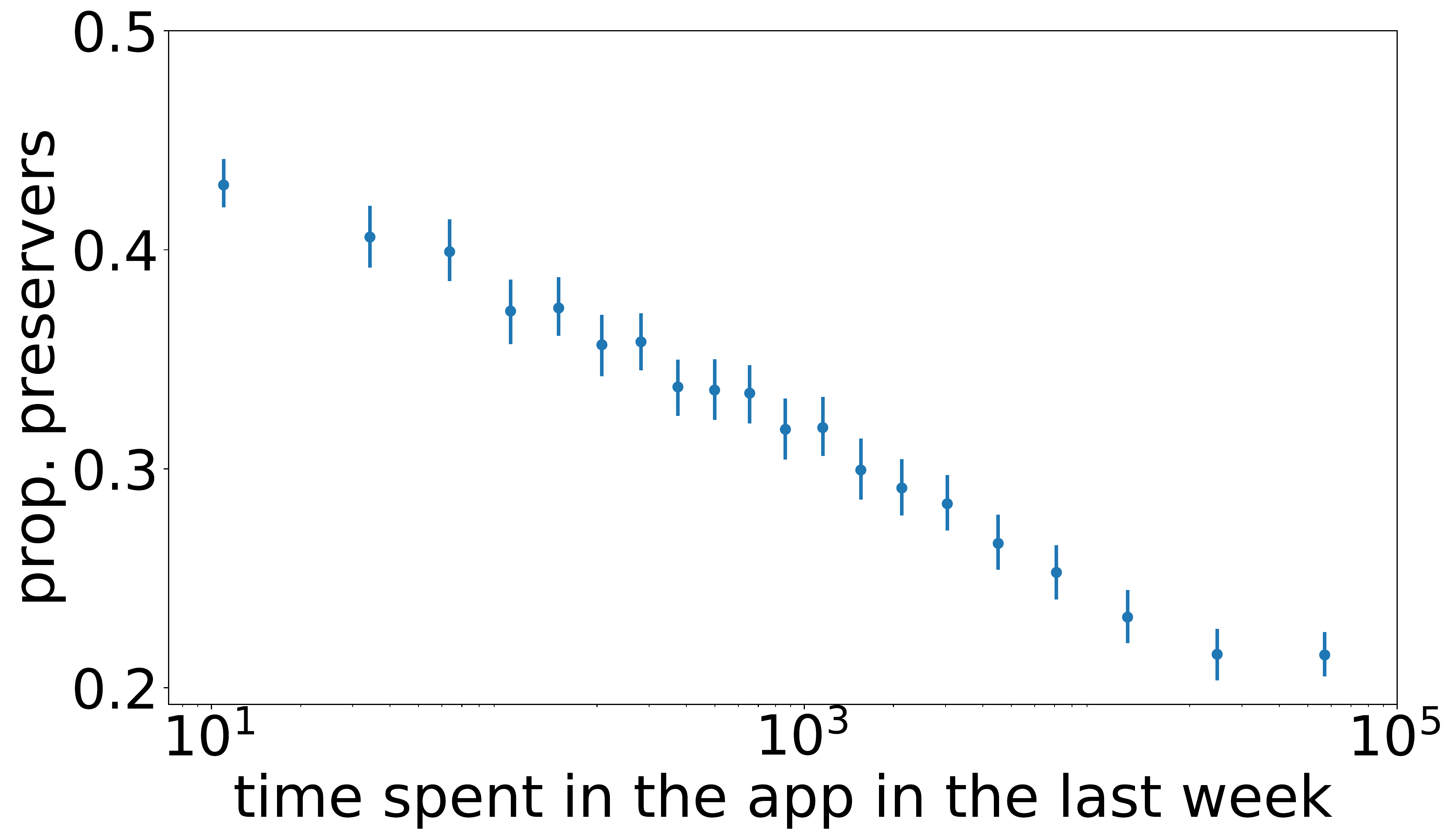}}
        \subfigure[Retro-adopters\label{fig:retro-adopters-last-week-duration}]
        {\includegraphics[width=0.24\linewidth]{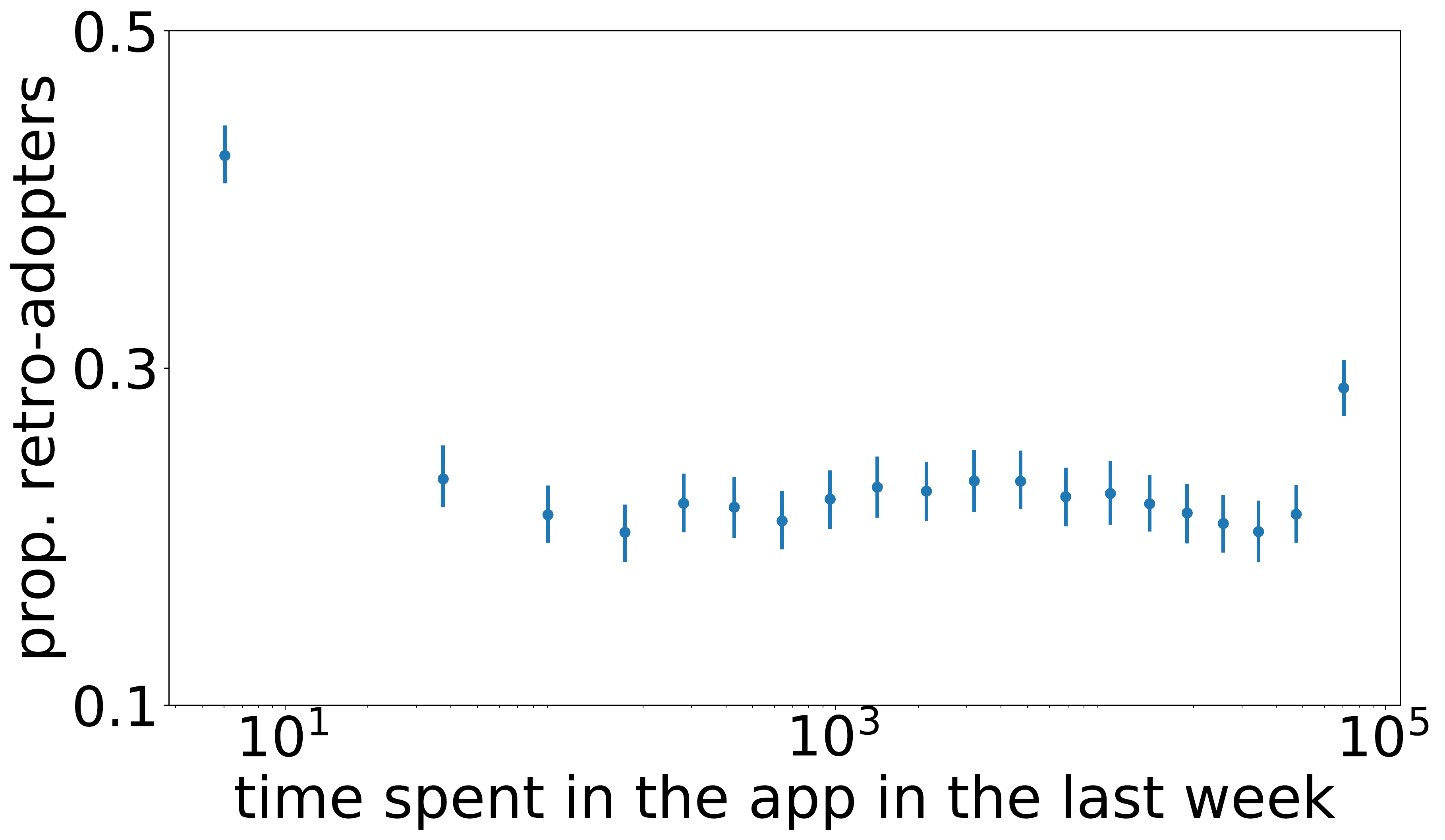}}
        \caption{Correlating adoption with previous usage.}\label{fig:last-week-duration}
    \end{center}
\end{figure*}


In the preceding prediction tasks, features describing users' interaction with the target app are selected as the most important ones in predicting ``adoption-or-not'' and ``time-to-adoption''. Motivated by this finding, we select typical features from this dimension, i.e., \textit{\#days to adopt the last version} and \textit{time spent in the app during the last week}, and correlate the adoption behaviors with them. Results are shown in Figure~\ref{fig:last-interval} and Figure~\ref{fig:last-week-duration}.

\noindent $\bullet$ \textbf{Correlation with previous adoption.} We first look at the relation between this feature and \#days to adopt the target version, which shows a positive correlation with a decelerate growth (see Figure~\ref{fig:all-users-last-interval}). The pattern in Figure~\ref{fig:subscribers-last-interval} complies with this finding, as users who take more days to adopt the previous version is less likely to be a subscriber of the new version, which means longer time to adopt it if they do. 
Figure~\ref{fig:dwellers-last-interval} shows that users who take more days to adopt the last version is more likely to be a preserver of the target version. This is quite reasonable as they could also take a long time to adopt the version next to the target version.
The bins in Figure~\ref{fig:retro-adopters-last-interval} show a V-shape pattern, indicating that users who take very short time or very long time to adopt the last version are more likely to be a retro-adopter. Such a pattern suggests that subscribers who adopt a new version fast could also return to a previous version decisively, while late adopters of previous versions could show low level of tolerance to changes.

\noindent $\bullet$ \textbf{Correlation with previous usage.} As for the second feature, we can also observe consistent patterns in its relation with the \#days to adopt (Figure~\ref{fig:all-users-last-week-duration}), the probability of being a subscriber (Figure~\ref{fig:subseribers-last-week-duration}), and the probability of being a preserver (Figure~\ref{fig:dwellers-last-week-duration}). It is understandable as users who spent more time in the app are more likely to know the update of the app in time, and take shorter time to adopt it. Figure~\ref{fig:retro-adopters-last-week-duration} shows that the probability of users being a retro-adopter stays stable with the change of usage during the previous week, with exceptions in both ends. It could be inferred from the pattern that heavy users of the previous version are with weak adaptability to the changes in the target new version, while the very light users might be unfamiliar with the app and prefer returning to previous versions. 

This analysis provides insights about why users behave as a subscriber, a preserver, or a retro-adopter. More importantly, the clear patterns shown in the correlations suggest that these special behaviors of adoption could be sensed and explained by our proposed features.

\section{Discussion}\label{sec:discussion}

Our study reveal novel findings about the adoption of recurrent innovations. We however recognize limitations of our analysis which readers should consider before applying our conclusions. 
First, because the data of adopting recurrent innovations are hard to obtain in other domains, we are unsure whether the results and findings from mobile app updates can be directly generalize to other domains. However, with the user behavior data we are able to have a first look at the adoption patterns of recurrent innovations, which should shed light on the analysis of other types of innovations when data become available.

Second, because the dataset is collected by a third-party business intelligence service provider, only apps that use their service are included. The features to characterize users with their interactions in other apps could be biased, which might be a reason for their insignificance in the prediction models.

Third, although the dataset is at a large scale to support our analysis, some important information are not collected or not revealed to us. Because the apps and versions are anonymized, we are not able to access the meta information of the apps and the description and content of each version. Thus, the difference of each app and each version could only be described with user behavior data. 

Other confounding factors that can not be derived from the dataset may affect the diffusion process of app versions. For example, users of different app markets (i.e., the communication channel in the diffusion theory) could differ in adopting an app version. The social network of users may have an impact on new adopters as well as recurrent adopters. Other characteristics of the users such as demographic information may also have a relation with their decision to adopt a recurrent innovation. Although we find similar patterns of adoption from the crowed for each version of an app, variations can still be observed and might be explained by such missed information. Unfortunately such user \& social network information is not available in our dataset.  We believe that with more information about the users and the innovations, the individual adoption behavior can be better understood. 

\section{Related Work}\label{sec:related}
In this paper we conduct the first systematic analysis of the diffusion of recurrent innovations with a large scale dataset of mobile app users consuming many versions of thousands of apps. In this section we compare our work with literature about adoption of mobile apps, and introduce related directions for further study of recurrent innovations with existing literature of innovation diffusion.

\subsection{Adoption of Mobile Apps}
Considerable effort in mobile app studies has been devoted to understanding the adoption of apps by mining the ratings, reviews, number of downloads, and other measures such as uninstalls~\cite{lu2017prado,li2016voting,DBLP:conf/icse/LuLLXMH0F16,DBLP:journals/tois/LiuALTHFM17,DBLP:conf/www/LiLAMF15,DBLP:conf/imc/LiLLXBLMF15} from app markets as indicators of users attitudes towards apps. Other research efforts focus on understanding the adoption of apps through in-app user behavior analysis, which including contextual data analysis~\cite{li2020systematic}, natural language data analysis~\cite{lu2016learning,ai2017untangling,chen2018through}, and so on~\cite{xu2017appholmes,shen2017towards,liu2019first,chen2020comprehensive,CHENDLDEPLOY2}. However, there are some limitations from the perspective of innovation diffusion. First, the object in these studies are usually the apps and the updates, if studied, are regarded as a feature of the apps. For example, Mcilroy et al.~\cite{mcilroy2016fresh} analyzed the updates for 10,713 Android apps and found that users highly ranked frequently-updated apps instead of being annoyed about the high update frequency. Second, the number of downloads or uninstalls are aggregated values and can not be traced back to individual adopters. Last but not least, the ratings and reviews are from part of adopters, which could cause selection bias. For example, Hassan et al.~\cite{hassan2017empirical} analyzed patterns of emergency updates of mobile apps and found a lower ratio of negative reviews.

Several pieces that study the adoption of app versions include the analysis of users' characteristics and their attitudes towards automatic mobile app updates~\cite{mathur2017impact}. This work conducted a survey with Android users and reveals that users who avoid auto-updates of apps are more likely to have had past negative experiences with software updating, tend to take fewer risks, and display greater proactive security awareness. Another work that worth mentioning analyzed how soon a user would update an app in Google Play~\cite{moller2012update}. They keep track of the installations of one app that they developed and published in Google Play over time, and find half of users of an old version did not update to a new version even 7 days after it is published. Their result also shows the effect from a new update on the diffusion of the previous update. Such findings complies with the curves of cumulative adopters in Figure~\ref{fig:total-line}. This work provided a first look at the adoption of recurrent innovations, although this term is not used by them and only one app is studied, and provides an evidence for our findings with app adopters using Google Play.

Another thread is to apply diffusion of innovation models to study the diffusion of specific apps in domains including health~\cite{murnane2015mobile, east2015mental}, traffic~\cite{yujuico2015considerations}, crisis~\cite{grinko2019adoption},  trekking~\cite{nickerson2014mobile}, etc., and to understand reasons for adoption such as personality traits~\cite{xu2016understanding} and life stage~\cite{frey2017mobile}. For example, Nickerson et al.~\cite{nickerson2014mobile} examined the diffusion of mobile technology and smartphone apps among people who walk the Camino de Santiago, a nearly 500 mile trek in Spain, with a research model that relates categories of adopter with the beliefs about innovation characteristics and with the adoption with innovations, respectively. They conclude that the model is only partially supported for this domain. Apps are regarded as single innovations and different versions are not considered as recurrent innovations.

\subsection{Diffusion of Recurrent Innovations}
Our work revels novel patterns of the adoption of recurrent innovations, which shed lights on the research of innovation diffusion. 
We next list the related research directions from the literature of innovation diffusion from where interested researchers can start with. 

\noindent $\bullet$ \textbf{Application to specific domains}. The diffusion of innovation researchers have applied the diffusion models in different disciplines such as medical sociology~\cite{coleman1957the}, cultural anthropology~\cite{barnett1963innovation}, industrial economics~\cite{mansfield1985rapidly} and health care~\cite{dearing2018diffusion}. With technology innovations updated recurrently, recurrent innovations could be observed in various fields ranging from automotive manufacturing to neural networks development. Exploring diffusion patterns of recurrent innovations in such domains would be crucial for different stakeholders and help polish the understanding of the nature of recurrent innovations, and enrich the theory. 

\noindent $\bullet$ \textbf{Communication channels}. Communication channels is a main element that influences the spread of an innovation~\cite{rogers2010diffusion}. In our case of mobile app updates, the communication channels could be the apps themselves as they can send notifications to mobile users, the app markets where users can be notified of new updates, and advertisements in various media. How the different channels and dissemination strategies applied in the channels influence the diffusion of the recurrent innovations need to be studied.

\noindent $\bullet$ \textbf{Characteristics of adopters}. The characteristics of adopters is also a main element of diffusion of innovation~\cite{rogers2010diffusion}. In this work we describe the adopters with their interactions with apps and their history versions. We find distinguished difference of new adopters and existing adopters and three categories of adopters with special adoption patterns. Can such patterns be explained by other characteristics of the adopters such as their demographics and personality? How to understand the innovativeness~\cite{rogers2010diffusion} of adopters towards recurrent innovations? Such questions need to be answered to portray the adopters of recurrent innovations more comprehensively.

\noindent $\bullet$ \textbf{Diffusion in social network}. The impact of social influence on the dynamics of diffusion has been extensively explored~\cite{valente1996social, katona2011network} from multiple aspects including the the local network structure, the characteristics of adopted neighbors, the distance to opinion leaders, etc. in the interpersonal networks. Such factors can also influence the diffusion of recurrent innovations.

\noindent $\bullet$ \textbf{Competition and collaboration of innovations}. Innovations could have competitors and collaborators. Taking the mobile apps as an example, apps with similar functions could be competitors of each other, while an app can collaborate with other apps by ways such as integrating a login authentication service provided by a social network app. Given that the interactions between innovations show effect in the diffusion of innovations~\cite{alon2010note, goyal2019competitive, tang2009social, weng2012competition, rong2013diffusion}, they could also influence the diffusion of recurrent innovations, which is worth of further study.

\section{Conclusion}\label{sec:conclusion}

In this paper, we present the first large-scale analysis of the adoption of recurrent innovations in the context of mobile app updates. Our analysis reveal novel patterns of crowd adopting behaviors with millions of users who consume the many versions of thousands of Android apps. We identify new categories of adopters to be added on the top of the Rogers' model of innovation diffusion, that is, subscribers, preservers, and retro-adopters. We show that standard machine learning models are able to predict users' decision of adopting a new version of an app by picking up various sources of signals from three groups of features, i.e., the properties of the technology, the characteristics of the adopter, and the how the adopter interacts with the recurrent innovations of the technology.

\begin{acks}

\end{acks}

\bibliographystyle{ACM-Reference-Format}
\bibliography{sample-base}

\appendix

\end{document}